\documentclass[useAMS,usenatbib]{mn2e}

\DeclareSymbolFont{cmletters}{OML}{cmm}{m}{it}
\DeclareMathSymbol{v}{\mathalpha}{cmletters}{"76}

\usepackage{tabularx,ragged2e,booktabs,caption}
\usepackage{amsmath}
\usepackage{amssymb}
\usepackage{epsfig}
\usepackage{graphicx}
\usepackage{ifthen}
\usepackage{latexsym}
\usepackage{rotating}
\usepackage{subfigure}
\usepackage{times,epsf}
\usepackage{txfonts}
\usepackage{varioref}
\usepackage{verbatim}
\usepackage{array}
\usepackage{url}
\usepackage{color}
\usepackage[dvipsnames]{xcolor}
\usepackage[T1]{fontenc}

\newcommand{\be}{\begin{equation}}
\newcommand{\ee}{\end{equation}}
\newcommand{\bea}{\begin{eqnarray}}
\newcommand{\eea}{\end{eqnarray}}
\newcommand{\gdet}{\sqrt{-g}}

\newcommand\apj{Astrophysical Journal}

\newcommand\apjs{Astrophysical Journal Suppl. Ser.}

\newcommand\aap{Astronomy \& Astrophysics}

\newcommand\prd{Physical Review D}
\newcommand\mnras{Monthly Notices of the Royal Astronomical Society}

\newcommand\pasj{Publications of the Astronomical Society of Japan}

\newcommand\jqsrt{Journal of Quantitative Spectroscopy and Radiative Transfer}

\newcommand{\pder}[2]{\frac{\partial#1}{\partial#2}}

\newcommand{\koral}{\texttt{KORAL}}
\newcommand{\Medd}{\dot M_{\rm Edd}}

\title[Global accretion disks with sub-grid dynamo]{
Global simulations of axisymmetric radiative black hole accretion disks in general relativity with a sub-grid magnetic dynamo}
\author[A. S\k{a}dowski, R. Narayan, A. Tchekhovskoy, D. Abarca,  Y. Zhu, J. C. McKinney]
       {Aleksander S\k{a}dowski$^1$\footnotemark[1], 
	Ramesh Narayan$^{1}$\footnotemark[1], 
        Alexander Tchekhovskoy$^{2}$\footnotemark[1],
        David Abarca$^{1}$\footnotemark[1],
        \newauthor
        Yucong Zhu$^{1}$\footnotemark[1]
        and Jonathan C. McKinney$^{3}$\thanks{E-mail: asadowski@cfa.harvard.edu (AS); 
rnarayan@cfa.harvard.edu (RN);	atchekho@princeton.edu (AT); dabarca@cfa.harvard.edu (DA); yzhu@cfa.harvard.edu (YZ); jmck@umd.edu (JM); } \\
        $^1$ Harvard-Smithsonian Center for Astrophysics, 60 Garden St., Cambridge, MA 02134, USA\\
        $^2$ Lawrence Berkeley National Laboratory, 1 Cyclotron Rd, Berkeley, CA 94720, USA; Einstein Fellow\\
        $^3$ University of Maryland at College Park, Dept. of Physics, Joint Space-Science Institute, 1117 John S. Toll Building \#082, College Park, MD 20742, USA }

\begin{document}

\maketitle

\label{firstpage}

\begin{abstract}

We present a sub-grid model that emulates the magnetic dynamo
operating in magnetized accretion disks. We have implemented this
model in the general relativisic radiation magnetohydrodynamic
(GRRMHD) code \koral, using results from local shearing sheet
simulations of the magnetorotational instability to fix the parameters
of the dynamo. With the inclusion of this dynamo, we are able to run
2D axisymmetric GRRMHD simulations of accretion disks for arbitrarily
long times.  The simulated disks exhibit sustained turbulence, with
the poloidal and toroidal magnetic field components driven towards a
state similar to that seen in 3D studies.  Using this dynamo code, we
present a set of long-duration global simulations of super-Eddington,
optically-thick disks around non-spinning and spinning black
holes. Super-Eddington disks around non-rotating black holes exhibit a
surprisingly large efficiency, $\eta\approx0.04$, independent of the
accretion rate, where we measure efficiency in terms of the total
energy output, both radiation and mechanical, flowing out to infinity.  
Super-Eddington disks around spinning black holes are even
more efficient, and appear to extract black hole rotational energy
through a process similar to the Blandford-Znajek mechanism.  All the
simulated models are characterized by highly super-Eddington radiative
fluxes collimated along the rotation axis.  We also present a set of
simulations that were designed to have Eddington or slightly
sub-Eddington accretion rates ($\dot{M} \lesssim 2\Medd$). None of
these models reached a steady state. Instead, the disks collapsed as a
result of runaway cooling, presumably because of a thermal
instability.

\end{abstract}


\section{Introduction}
\label{introduction}

Black hole (BH) accretion disks are involved in some
of the most energetic phenomena in the Universe - active galactic
nuclei (AGN), microquasars, tidal disruptions of stars, 
$\gamma$-ray bursts. They power relativistic jets and
provide feedback on the large-scale evolution
of galaxies. Understanding the structure of accretion disks is therefore
crucial for many applications in astrophysics.

Numerical simulations are a powerful tool for studying
BH accretion disks, and can fill the gap
between observational data (which are often limited 
because of the large distance to most objects),
and theory (which is incapable of fully describing the non-linear
turbulence in disks).

Gas can be accreted on BHs through hot or cold accretion flows
\citep{yuannarayan-14}.  The former are optically thin and
predominantly radiatively inefficient because of long cooling
timescales.  The latter are optically thick, and may be either
radiatively efficient (standard thin disks), or inefficient
(slim disks) because of long radiative diffusion timescales.
Optically thin flows have been extensively studied by means 
of general relativistic (GR) magnetohydrodynamical (MHD)
simulations 
\citep[e.g.,][]{devilliersetal03,gammie03,anninosetal05,delzannaetal07,narayan+12,mtb12,tchekh+12}.
In optically thin flows, the radiation field is virtually decoupled from
the gas, allowing for the radiation field to be neglected. 
On the other hand, simulating optically thick accretion is far more 
demanding since it requires coevolution of both the radiation and gas fields.

Numerical simulations of systems with radiation
is a non-trivial problem; Solving the radiation transport equations
is often overwhelmingly expensive, so various approximations are needed.
The most common approach in the past was to invoke flux limited diffusion (FLD)
\citep[e.g.,][]{hirose09a,blaes2011,ohsuga09,ohsuga11},
which assumes that the radiative flux follows the gradient of the radiative energy 
density. Its Achilles heel is that it does not satisfy momentum conservation (since FLD does not keep track of the radiation pressure) and also does not work properly in the optically thin limit.

As an alternative to FLD, the M1 radiative closure \citep{levermore84,dubrocafeugeas99}
has recently been adopted by a number of groups \citep{sadowski+koral,skinnerostriker-13,mckinney+harmrad} as a more accurate treatment of the radiation field.  The advantage of M1 is that it addresses the two fundamental problems of FLD (momentum
conservation and optically thin limit) while also working in a general relativistic (GR) framework. However, M1
has its own limitations -- it cannot handle
photon distributions with complex angular structure. \cite{jiang+12} have overcome the latter limitation by 
estimating the radiative stress energy tensor through a short-characteristics
solver that makes use of a large number of angles. They have applied the code
to shearing sheet local simulations \citep{jiang+13}. Very recently, \cite{jiang+14}
have switched to an even more self-consistent approach where specific intensities at fixed angles are evolved directly, and the radiative
stress energy tensor is constructed on the go from the exact intensity distribution.

Although many numerical codes are available for evolving 
black hole accretion disks with gas and radiation (see above),
even the simplest problems are still prohibitively expensive to simulate in full 3D.
One saving grace is the case of very large accretion rates (i.e. exceeding the Eddington accretion
rate\footnote{In this work we adopt the following definition
for the Eddington accretion rate,
\be
\label{e.medd}
\Medd = \frac{L_{\rm Edd}}{\eta_0 c^2},
\ee
where $L_{\rm Edd}=1.25 M/M_{\odot}\times 10^{38} \rm ergs/s$ is the 
Eddington luminosity, and $\eta_0$ is the radiative efficiency of a thin
disk for a given spin $a_*$,
\be
\label{e.eta}
\eta_0= 1-\sqrt{1-\frac{2}{3\,R_{\rm ISCO}}},
\ee
and $R_{\rm ISCO}$ is the radius of the Innermost Stable Circular Orbit (ISCO).
 According to this definition,
any thin, radiatively efficient disk accreting at $\Medd$ has luminosity $L_{\rm Edd}$.
}).  Highly accreting systems are feasible to simulate in three dimensions \citep{mckinney+harmrad} because their short
viscous timescales and large geometrical thickness allows for relatively short duration of simulations and low resolution, respectively.
Simulating disks at lower accretion rates is much more 
demanding and may be beyond reach for the near future.

Obviously, axisymmetric, two-dimensional (2D) simulations are
computationally much less expensive than three-dimensional (3D), so
one can simulate a wider range of systems in 2D. What has prevented
such simulations in the past is the anti-dynamo theorem
\citep{cowling-33}, which states that axisymmetric ideal MHD systems do not have
any dynamo mechanism to regenerate the magnetic field This means 2D
axisymmetric simulations cannot regenerate their magnetic fields and
therefore accretion will die after a short time --- i.e. the poloidal magnetic
field quickly decays and it is not replaced because of this lack of
dynamo.  To get over this problem, we introduce here a sub-grid dynamo
model which emulates the missing dynamo process, and drives the
properties of turbulence towards a state characteristic of disk MRI
turbulence in local 3D simulations. We apply our method to radiative
accretion flows and simulate a number of disks models covering a wide
range of accretion rates.

The paper is organized as follows. In Section \ref{s.methods}
we briefly introduce the numerical methods we use, with details
given in Appendices \ref{ap.dynamo} and \ref{ap.viscosity}.
In Section \ref{s.simulations} we present the global simulations
we performed. Finally, in Secion \ref{s.summary} we summarize
our work.

\section{Numerical methods}
\label{s.methods}

\subsection{GRRMHD code --- \texttt{KORAL}}
\label{s.koral}

We use the general relativistic radiation magnetohydrodynamics (GRRMHD) code
\texttt{KORAL} \citep{sadowski+koral, sadowski+koral2}, which employs a Godunov scheme
to evolve the rest mass, energy and momentum conservation equations in
a fixed, arbitrary spacetime using finite-difference methods. 
The magnetic field is evolved according to the induction equation and 
the divergence-free criterion is enforced using the flux-constrained \cite{toth-00}
scheme as described in \cite{gammie03}. \texttt{KORAL} simultaneously
evolves two fluids, magnetized gas and radiation, which
exchange energy and momentum via mutual interactions.

The conservation laws are described by the following general set of 
equations,
\bea\label{eq.rhocons}
\hspace{1in}(\rho u^\mu)_{;\mu}&=&0,\\\label{eq.tmunucons}
\hspace{1in}(T^\mu_\nu)_{;\mu}&=&G_\nu,\\\label{eq.rmunucons}
\hspace{1in}(R^\mu_\nu)_{;\mu}&=&-G_\nu,
\eea
where $\rho$ is the gas
density in the comoving fluid frame, $u^\mu$ is the gas four-velocity
as measured in the ``lab frame'', and $T^\mu_\nu$ is the
MHD stress-energy tensor in this frame,
\be\label{eq.tmunu}
T^\mu_\nu = (\rho+u_{\rm g}+p_{\rm g}+b^2)u^\mu u_\nu + (p_{\rm g}+\frac12b^2)\delta^\mu_\nu-b^\mu b_\nu,
\ee 
$R^\mu_\nu$ is the stress-energy tensor of radiation, and $G_\nu$ is the radiative
four-force describing the interaction between gas and radiation (both are described in detail in Section~\ref{s.m1}). Here $u_{\rm g}$ and $p_{\rm g}=(\Gamma-1)u_{\rm g}$ represent the internal energy and pressure of the 
gas in the comoving frame and $b^\mu$ is the magnetic field 4-vector \citep{gammie03}.
The magnetic pressure is $p_{\rm mag}=b^2/2$.

In the coordinate basis Eqs.~(\ref{eq.rhocons})-(\ref{eq.rmunucons}) take the form,
\bea\label{eq.cons3_1}
\hspace{.3in}\partial_t(\gdet\rho u^t)+\partial_i(\gdet\rho u^i)&=&0,\\\label{eq.cons3_2}
\hspace{.3in}\partial_t(\gdet T^t_\nu)+\partial_i(\gdet T^i_\nu)&=&\gdet T^\kappa_\lambda \Gamma^\lambda_{\,\,\nu\kappa} + \gdet G_\nu,\\\label{eq.cons3_3}
\hspace{.3in}\partial_t(\gdet R^t_\nu)+\partial_i(\gdet R^i_\nu)&=&\gdet R^\kappa_\lambda \Gamma^\lambda_{\,\,\nu\kappa} - \gdet G_\nu,
\eea
where $\gdet$ is the metric determinant, and $\Gamma^\lambda_{\,\,\nu\kappa}$ are Christoffel symbols.

\texttt{KORAL} adopts the ideal MHD approximation and assumes that the electric
field vanishes in the fluid rest frame. The induction equation 
then takes the following form in coordinate basis,
\be
\label{eq.Maxi}
\partial_t(\sqrt{-g}B^i)=-\partial_j\left(\sqrt{-g}(b^ju^i-b^iu^j)\right),
\ee
where $B^i$ is the magnetic field three-vector \citep{komissarov-99} which satisfies,
\be
\label{eq.Bit}
b^t=B^i u^\mu g_{i\mu},
\ee
\be
\label{eq.Bi1}
b^i=\frac{B^i+b^tu^i}{u^t}.
\ee 
The  
flux-interpolated contrained transport (Flux-CT) method of \cite{toth-00}
prevents numerical generation of spurious magnetic monopoles.
For the gas and magnetic fields, we use
standard numerical methods as described in previous papers 
\citep{gammie03,mckinney06,mtb12,sadowski+koral, sadowski+koral2}.

\subsection{Radiative closure}
\label{s.m1}

At each time step, radiative energy density ($R_t^t$) and radiative
fluxes ($R_i^t$) are evolved following Eq.~(\ref{eq.cons3_3}).
To calculate the time derivatives we need to know all the 
remaining components of 
radiation stress-energy
tensor $R_\nu^\mu$.
For this purpose we make use of
the M1 closure scheme \citep{levermore84}. In this approach, we assume that the radiation
tensor is isotropic and satisfies Eddington closure, not in the
fluid frame, but in the orthonormal ``rest frame'' of the
radiation. The latter is defined as the frame in which the radiative
flux vanishes. 

A covariant formalation of the M1 scheme was introduced 
in \cite{sadowski+koral} which we have adopted. Herein, we give only the essential formulae and ask the reader to refer to
that paper for details.

Knowing $R^{t\mu}$ we calculate the time-component of the radiative rest-frame
four velocity, $u^t_R$, and the radiative energy density in this frame, $E_R$, by solving
the following two equations,
\be
g_{\mu\nu}\,R^{t\mu}R^{t\nu} = -\frac{8}{9}E_R^2 (u^t_R)^2
+\frac{1}{9}E_R^2 g^{tt},
\ee
\be
R^{tt} = \frac{4}{3}E_R (u^t_R)^2 + \frac{1}{3}E_R g^{tt}.
\label{eq:invert2}
\ee
These quantities are then used to find the spatial components of $u^\mu_R$
using the time component of,
\begin{equation} 
R^{\mu\nu} = \frac{4}{3} E_R\, u^\mu_R u^\nu_R + \frac{1}{3}E_R\,
g^{\mu\nu}.
\label{eq:R}
\end{equation}
Once we have $u^\mu_R$ and $E_R$, we use the remaining components of Eq.~(\ref{eq:R}) to compute
full radiative stress energy tensor
$R^{\mu\nu}$.

M1 closure is superior to the Eddington closure and FLD, especially for
optically thin media. It is a simple and elegant closure scheme
which evolves both the radiation energy density and radiation fluxes.
However, M1 allows for only a very limited set of the specific intensity distributions 
(boosted isotropic), and therefore is far from perfect. 
More advanced radiation transfer schemes have been developed \citep[e.g.,][]{jiang+12,jiang+14}.
The main advantage of M1 closure over these more advanced schemes
is that it is covariant, fast, and local; also, it has been implemented
in radiation GRMHD codes whereas none of the other schemes have yet been
attempted with general relativity. In our view, it is reasonable to assume
that M1 catches most of the relevant dynamics of radiative
BH accretion flows.

\subsection{Radiative four-force}

The four-vector $G^\mu$ describes the interaction between gas and radiation.
\cite{sadowski+koral2} introduced covariant formalism for computing the interaction
due to absorption and elastic scattering.
In this paper, we account also for energy exchange via Comptonization,
which we implement in a similar way to that described in \cite{kawashima+09}. Detailed derivation is 
given below in Section~\ref{s.comptonization}.

We write the four-force $G^\mu$ as
\be
\label{eq.Gmutotal}
G^\mu=G^\mu_0 + G^\mu_{\rm Compt},
\ee
where 
$G^\mu_{\rm Compt}$ describes the effect of Comptonization, and $G^\mu_0$ corresponds to absorption and Thomson scattering
and is given by \citep{sadowski+koral2},
\be
\label{eq.Gcon}
 G_0^\mu = -\rho (\kappa_{\rm a}+\kappa_{\rm es}) R^{\mu\nu}  u_\nu
-\rho \left(\kappa_{\rm es} R^{\alpha\beta}  u_\alpha  u_\beta+\kappa_{\rm a} 4\pi  B\right) u^\mu,
\ee
where $\kappa_{\rm a}$ and $\kappa_{\rm es}$ are absorption and scattering opacities,
respectively, and $B=aT_{\rm g}/4\pi$ ($a$ is the radiation constant) is the intensity of black body radiation 
for gas with temperature $T_{\rm g}$.

The source term associated with the radiative four-force is stiff 
whenever the gas is optically thick.
\texttt{KORAL} deals with this problem by applying it
semi-implicitly. That is, at each cell center, after applying the advective operator (the spatial gradient terms in equations \ref{eq.cons3_1}-\ref{eq.cons3_3}), the following set of equations is
solved,
\bea
&&\hspace{1cm}T^t_{\nu,(n+1)}-T^t_{\nu,(n)}=\Delta t ~G_{\nu,(n+1)},\label{eq.source3}\\
&&\hspace{1cm}R^t_{\nu,(n+1)}-R^t_{\nu,(n)}= - \Delta t~ G_{\nu,(n+1)}, \label{eq.source4}
\eea
where the subscripts $(n)$ and $(n+1)$ denote values at the beginning and end of a 
time step of length $\Delta t$, respectively. For details of this procedure, see \cite{sadowski+koral2}.

\subsection{Thermal Comptonization}
\label{s.comptonization}

In the absence of Comptonization, in fluid frame
orthonormal coordinates the time and spatial components of the
radiative four-force (Eq.~\ref{eq.Gcon}) take the form,
\bea
\hspace{1.in}\widehat{G}^0 &=& \kappa_{\rm a}\rho (\widehat{E}-4\pi \widehat{B}), \label{G0}
\\
\hspace{1.in}\widehat{G}^i &=& (\kappa_{\rm a}+\kappa_{\rm es}) \rho \widehat{F}^i,
\label{Gi}
\eea
where
\begin{equation}
4\pi \widehat{B} = aT_{\rm g}^4, \label{Tg}
\end{equation}
and $T_{\rm g}$ is the temperature of the gas. 
Equation (\ref{G0}) describes the rate of change of the fluid energy
density as a result of energy gain through absorption, $\kappa_{\rm
  a}\rho\widehat{E}$, and energy loss through emission, $\kappa_{\rm
  a}\rho (4\pi \widehat{B})$. For simplicity,
we treat radiation as a blackbody, hence
we define an effective radiation temperature $T_{\rm r}$ in the
fluid frame via
\begin{equation}
\widehat{E} = aT_{\rm r}^4. \label{Tr1}
\end{equation}
Thus, equation (\ref{G0}) shows that gas gains energy at a rate
proportional to $T_{\rm r}^4$ and loses energy proportional to $T_{\rm
  g}^4$. As a result, the two temperatures are pushed towards each
other, i.e., the system is driven towards thermal equilibrium. Note
that $\kappa_{\rm es}$ does not appear in the energy equation. This is
because Thomson scattering redirects photons, but it does not transfer
energy.

Equation (\ref{Gi}) describes the rate of change of the fluid momentum
density. The gas acquires the momentum of each photon that it either
absorbs or scatters. Since the re-emission of radiation
is symmetric in the fluid
frame, there is no  counter-balancing term with a negative sign.  Note
that the fluid gains momentum density in a direction parallel to the
radiation flux $\hat{F}^i$, and the radiation loses a corresponding
amount of momentum density. Hence the system is driven towards a state
in which there is no relative motion between the fluid and radiation
frames, i.e., no radiation flux in the fluid frame ($u^\mu=u_R^\mu$).

The main effect of Comptonization is that scattering causes not just
momentum transfer between the radiation and gas, but also energy
transfer. A soft photon of energy $\epsilon_0$ which scatters off a
thermal electron with temperature $T_{\rm e}$ on average gains an
energy $\langle\Delta\epsilon\rangle$ given by the following two
expressions in the non-relativistic and ultra-relativistic limits:
\begin{eqnarray}
\langle\Delta\epsilon\rangle &=& \left(\frac{4kT_{\rm e}}{m_{\rm e}c^2}\right)\epsilon_0,
\qquad 4kT_{\rm e} \ll m_{\rm e}c^2, \label{CompNR} \\
\langle\Delta\epsilon\rangle &=& \left(\frac{4kT_{\rm e}}{m_{\rm e}c^2}\right)^2\epsilon_0,
\qquad 4kT_{\rm e} \gg m_{\rm e}c^2. \label{CompR}
\end{eqnarray}
For a general temperature, using the result given in equation (2.43) in
\cite{pozdnyakov+83}, we have obtained a good fitting
function (maximum fractional error 1.2\%) which works for all $T_e$
\begin{equation}
\langle\Delta\epsilon\rangle = \epsilon_0 
\left(\frac{4kT_{\rm e}}{m_{\rm e}c^2}\right) 
\left[1+3.683 \left(\frac{kT_{\rm e}}{m_{\rm e}c^2}\right) 
+4 \left(\frac{kT_{\rm e}}{m_{\rm e}c^2}\right)^2\right]
\left[1+\left(\frac{kT_{\rm e}}{m_{\rm e}c^2}\right)\right]^{-1}.
\label{Compgen} \\
\end{equation}

The above expressions are valid so long as the photon is soft, i.e.,
the radiation temperature is much less than the gas temperature. When
the two temperatures are equal, thermodynamics guarantees that there
is no energy transfer between gas and radiation. Similarly, when the
radiation temperature is larger than the gas temperature, we expect
energy to flow from the radiation to the gas. To allow for these
effects, we modify equation (\ref{G0}) by introducing an extra contribution to 
$\widehat{G}^0$,
\bea
\label{CompG0}
\widehat{G}^0_{\rm Compt} &=& -
\kappa_{\rm es}\rho\widehat{E} \left[\frac{4k(T_{\rm g}-T_{\rm
      r})}{m_{\rm e}c^2}\right] \times\\\nonumber
&&\times\left[1+3.683 \left(\frac{kT_{\rm e}}{m_{\rm e}c^2}\right) 
+4 \left(\frac{kT_{\rm e}}{m_{\rm e}c^2}\right)^2\right]
\left[1+\left(\frac{kT_{\rm e}}{m_{\rm e}c^2}\right)\right]^{-1},
\eea
where we have replaced $T_e$ by $T_{\rm g}$.
The negative sign in the Compton term is because gas cools when
$T_{\rm g} > T_{\rm r}$. The cooling is proportional to the radiation
energy density, $\widehat{E}$, and to the number of scatterings per
unit time, $\kappa_{\rm es}\rho$. Except for the two final factors in square
parentheses, which are an approximate correction for relativistic
temperatures, equation (\ref{CompG0}) is
identical to the prescription used by \cite{kawashima+09}.

As far as the momentum equation is concerned, we assume that the
Compton-scattered radiation is symmetric in the fluid frame and
carries no net momentum (a fairly good approximation in the soft
photon limit).  Under this approximation, we do not
modify equation (\ref{Gi}).

Eq.~(\ref{CompG0}) gives the fluid-frame energy transfer rate due
to Comptonization. To obtain the corresponding ``lab frame'' four-vector
$G^\mu_{\rm Compt}$ (Eq.~\ref{eq.Gmutotal}) we write,
\be
G^\mu_{\rm Compt}=\widehat{G}^0_{\rm Compt}u^\mu.
\ee

We have included the above version of Comptonization in the
simulations described in this paper. That is, we solve equations
(\ref{eq.tmunucons}) and (\ref{eq.rmunucons}) using equations
(\ref{CompG0}) and (\ref{Gi}), where $\widehat{B}$ is given by
equation (\ref{Tg}) and $T_{\rm r}$ is given by equation
(\ref{Tr1}). Note that this is an extremely simple prescription. The
main weakness of this approach is that it does not conserve photon
number during scattering. Instead it assumes perfect blackbody and
uses equation (\ref{Tr1}) to obtain the radiation temperature, thus
missing any effects associated with spectral hardening and a dilute
blackbody.

\subsection{Sub-grid dynamo}
\label{s.dynamo}

Magnetic stresses generated via MRI turbulence are responsible for
angular momentum transport and energy dissipation in BH accretion
disks. The same turbulence also dissipates magnetic field. The field
is,
however, regenerated through a dynamo
\citep[e.g.,][]{parker-55,brandenburg+95}. The balance of these
two processes leads to a saturated quasi-equilibrium state.

Evolution of magnetic fields has been studied extensively by means
of local shearing box simulations 
\citep[e.g.,][]{turner2003,kro07,blaes2007,blackman+08,guan+09},
as well as global simulations \citep{sorathia+12}. It has been shown that
for the Keplerian shear the magnetic field saturates at a turbulent state
characterized by a mean magnetic field angle,
 \be
\label{eq.bangle}
\xi =\frac{ \hat b^r \hat b^\varphi
}{b^2}\approx 0.25,
\ee
and at magnetic to total pressure ratio,
 \be
\label{e.betaprime}
\beta' = \frac{p_{\rm mag}}{p_{\rm tot}}\approx0.1.
\ee

Reaching the saturated quasi-stationary state is possible only
in 3D simulations. If only axisymmetry is assumed,
the anti-dynamo theorem \citep{cowling-33} implies that the
magnetic field cannot be maintained by dynamo action. 
Axisymmetric MHD simulations of accretion disks are therefore not very 
useful. Not only is their duration limited (decaying poloidal magnetic field
implies decaying turbulence), but even at early times in these
simulations the configuration of magnetic field is far
from that expected in the saturated state of 3D MRI turbulence. Typically, an axisymmetric simulation
may last up to $\sim 5000 GM/c^2$, i.e., 5 orbits at radius $R=20$.

The evolution of the mean magnetic field may be described 
under the mean field theory by the following equation \citep{brandenburg-01},
\be
\label{eq.meandynamo}
\frac\partial{\partial t}\vec B=\alpha \nabla \times \vec B + \eta \nabla^2 \vec B,
\ee
where $\alpha$ and $\eta$, are the dynamo, and magnetic diffusivity 
coefficients, respectively. The first term on the right hand
side describes the dynamo effect which generates magnetic field, and
the second corresponds to the dissipation of magnetic field.

Direct implementation of the mean field equation into axisymmetric 
models is possible in resistive, non-ideal GRMHD codes 
\citep{buccidelzanna+13}. However, such dynamo closure is not unique,
and one still has to arbitrarily (basing on three dimensional studies)
specify values of $\alpha$ and $\eta$.

In this paper we present a simpler method, which could be used
in non-resistive, ideal MHD codes, i.e., codes which assume that the
electric field disappears in the gas comoving frame. 
We use Eq.~\ref{eq.meandynamo} and the expected 
properties of the magnetic field in the saturated state (Eqs.~\ref{eq.bangle} - \ref{e.betaprime}), and
construct a sub-grid dynamo model which injects a weak poloidal magnetic
field into the simulation on top of the preexisting magnetic field, and 
drives the total field towards the prescribed characteristics
consistent with the saturated state. A detailed derivation is given in Appendix~\ref{ap.dynamo}. Below
we give only the ultimate formulae implemented in \texttt{KORAL}. 

At each time step, an estimate of the dynamo-generated magnetic field
is superimposed on the existing magnetic field. The change in the poloidal 
magnetic field is calculated through the toroidal vector potential,
\be
\label{eq.dynamoaphi}
dA_\varphi = -\alpha_{\rm dyn} \, \Omega_{\rm K}
Rg_{\varphi\varphi}\, B^\varphi f_R f_\theta f_\xi f_{\rm eq}\,dt,
\ee
where $\alpha_{\rm dyn}$ is an arbitrary coefficient, $dt$ is the time step,
 $\Omega_{\rm K}=R^{-3/2}$, $\theta$ is the polar 
coordinate, $B^\varphi$ is the toroidal component of the magnetic
three-vector (Eqs.~\ref{eq.Bit}-\ref{eq.Bi1}), $f_R$, $f_\theta$ and
$f_\xi$ are arbitrary factors damping the dynamo process 
in the plunging region, outside the disk, and in regions with too large
magnetic field angle $\xi$, respectively, and $f_{\rm eq}$ makes the dynamo 
flip sign accross the equatorial plane. The dynamo-generated vector 
potential $dA_\varphi$ is then converted into poloidal magnetic field
through $\vec {dB}=\nabla \times \vec {dA}$, which is then superimposed
on top of the existing magnetic field.

Shear constantly generates azimuthal magnetic 
field from radial field. In three dimensions, the field strength saturates because of dissipation.
We mimic this saturation by damping the azimuthal component of
the lab-frame magnetic field in regions where magnetic pressure 
exceeds the prescribed pressure ratio $\beta'$. We do it on the
orbital timescale according to,
\be
\label{eq.dynamodamp}
dB^\varphi=-\alpha_{\rm damp}\, \Omega_{\rm K} B^\varphi f_R f_\theta f_{\beta'}\,dt,
\ee
where $\alpha_{\rm damp}=1.0$, and $f_{\beta'}$ is a factor
which switches off the damping in regions of too low pressure ratio $\beta'$.      
Because of axisymmetry, changing $B^\varphi$ does not violate the
divergence-free condition.
Detailed formulae are
given in Appendix~\ref{ap.dynamo}.

The sub-grid dynamo model is applied at every time substep,
after all the other operators have been accounted for. First, the vector
potential $dA_\varphi$ (Eq.~\ref{eq.dynamoaphi}) is converted to poloidal magnetic
$dB^p_{\rm dyn}$, and is added to the existing field $B^p$.  Secondly, the azimuthal magnetic field is damped according
to Eq.~\ref{eq.dynamodamp}. The total MHD 
stress-energy tensor components $T^\mu_t$ (the conserved quantities, Eq.~\ref{eq.tmunu})
are left unchanged during both steps. Thus, any change in the magnetic
energy density causes a compensating change in the gas internal energy.

\subsection{Radiative viscosity}
\label{s.viscosity}

In this work, we slightly modify the radiative 
flux terms in Eq. \ref{eq.cons3_3} by introducing a radiative viscosity
which is effective only in the optically thin region
and which helps to avoid artificial centrifugal 
shocks, which are known to occur near the polar axis with
M1 closure \citep{sadowski+koral}. We replace
the original, M1-based flux term, $R^i_{\nu,\rm M1}$, with,
\be
R^i_\nu=R^i_{\nu,\rm M1}+R^i_{\nu,\rm visc},
\ee
where $R^i_{\nu,\rm visc}=-2\nu E_{R} \sigma^{i}_\nu$ is the viscous
correction to the radiative stress-energy tensor, which depends
on the photon mean-free path, radiative energy density in the
radiation
rest frame, $E_R$, and the shear tensor, $\sigma^{i}_\nu$, estimated
from the velocities of the radiative rest frames of the given cell and its
neighbours.

Detailed formulae, and tests are given in Appendix \ref{ap.viscosity}.

\section{Global simulations}
\label{s.simulations}

\subsection{Numerical setup}
\label{s.numerical setup}

All the simulations were performed in 2.5 dimensions, i.e., we assumed
axisymmetry, but we allowed for non-zero $\varphi$-components of vectors
(e.g., angular velocity and azimuthal radiative flux). We used Kerr-Schild
horizon-penetrating coordinates. The internal grid ($x_1,x_2$) was uniform and was
related to the Kerr-Schild radial and polar coordinates ($R,\theta$) by,
\bea
\hspace{1cm}&&R=R_0+e^{x_1},\\
\hspace{1cm}&&\theta=\left(\frac{\tan\, (H_0 \pi (x_2 - 0.5))}{\tan\,(H_0 \pi/2 )}+1\right)\frac\pi 2.
\eea
As a result, the grid points were spaced roughly logarithmicaly in radius, and concentrated towards
the equatorial plane, with the density of the points there depending on $H_0$. 
The internal radial coordinate $x_1$ was chosen to correspond to the range of radii between 
$R_{\rm min}$ and $R_{\rm max}$, while $x_2$ covered range $(\epsilon,\, 1-\epsilon)$, with $\epsilon=0.005$.
Values of $R_{\rm min}$, $R_{\rm max}$, $R_0$, and $H_0$,  for each simulation are given in
Table~\ref{t.models}.

At the inner radial boundary (located at least five cells under the BH horizon) we applied 
an outflow boundary condition by copying values of all the primitive 
quantities to the ghost cells. At the outer radial boundary we adopted
a similar approach, except that
we prevented the inflow of gas or radiation by resetting
negative radial velocities to zero. The polar axis was treated with reflective boundary condition. To ensure stability in this region
the primitive quantities in the two closest cells to the polar axis were
appropriately overwritten with the values from the third
cell \citep{mtb12}. Scalars, radial and azimuthal components of vectors were copied, and the polar
components were interpolated towards zero to satisfy this reflective
boundary condition. As described in \cite{sadowski+koral},
we evolved entropy as an auxiliary quantity which was used when the regular,
energy-based, inversion failed. 

We adopted the sub-grid dynamo prescription (Section~\ref{s.dynamo} and Appendix~\ref{ap.dynamo}) which 
mimics the three-dimensional evolution of magnetic fields. 
For all the runs we adopted $\alpha_{\rm dyn}=0.05$,
$\alpha_{\rm damp}=1.0$, $\xi_{\rm dyn}=0.25$, and $\beta'_{\rm damp}=0.1$.
We also included the radiative viscosity 
(Appendix~\ref{ap.viscosity}) with $\alpha_{\rm rad}=0.1$.

\subsection{Initial state}
\label{s.initial}

All the simulations were started from an equilibrium torus rotating around
a $10 M_\odot$ BH, which was threaded by initial, poloidal seed magnetic field.

The gas was initially set as a hydrodynamical equilibrium torus following \cite{penna-limotorus}. Its angular velocity
at the equatorial plane 
was set to a constant fraction of $\xi=0.708$ of the Keplerian angular velocity
outside radius $R_1=42$, and followed fixed angular momentum inside that radius.
The angular momentum was kept constant along the von-Zeipel cylinders. The inner edge
of the torii was chosen at $R_{\rm in}=22$. The density was fixed by setting the 
entropy constant $\cal{K}$ to the values given in Table~\ref{t.models}, and by assuming $\Gamma=5/3$. 
This setup results in an equilibrium torus that has a
vertical surface density profile that is nearly constant with radius. A radius independent density profile was desired since it 
helps keep the accretion rate constant with time. On the other hand, the radial extent of the torus was limited
only to $R\lesssim 400$, and this fact could limit the available
range of the inflow equilibrium.

So far, we have specified a torus in pure hydrodynamical equilibrium.  To
introduce the initial radiation field, we decided to start simulations from
the local thermal equilibrium (LTE) by solving for gas (and radiation)
equilibrium temperature $T$, using, \be p_{\rm tot}=p_{\rm gas}+p_{\rm
  rad}=k_{\rm B}\rho T + \frac 43 \sigma T^4, \ee where $p_{\rm tot}$
is the total pressure given by the hydrodynamical torus solution, and
$p_{\rm gas}$ and $p_{\rm rad}$ are gas and radiation pressures of the
LTE torus, respectively.  Because of the inconsistency in the values
of $\Gamma$ of the initial hydrodynamical torus ($\Gamma=5/3$),
and the resulting effective $\Gamma$ of the gas and radiation mixture
($\Gamma\approx 4/3$ for radiation pressure dominated medium),
the torus is not in the perfect equilibrium at early times. However,
it remains in a relatively steady state until the magnetorotational
instability (MRI) grows, after which the initial state is forgotten.

\begin{figure}
\includegraphics[width=1.075\columnwidth]{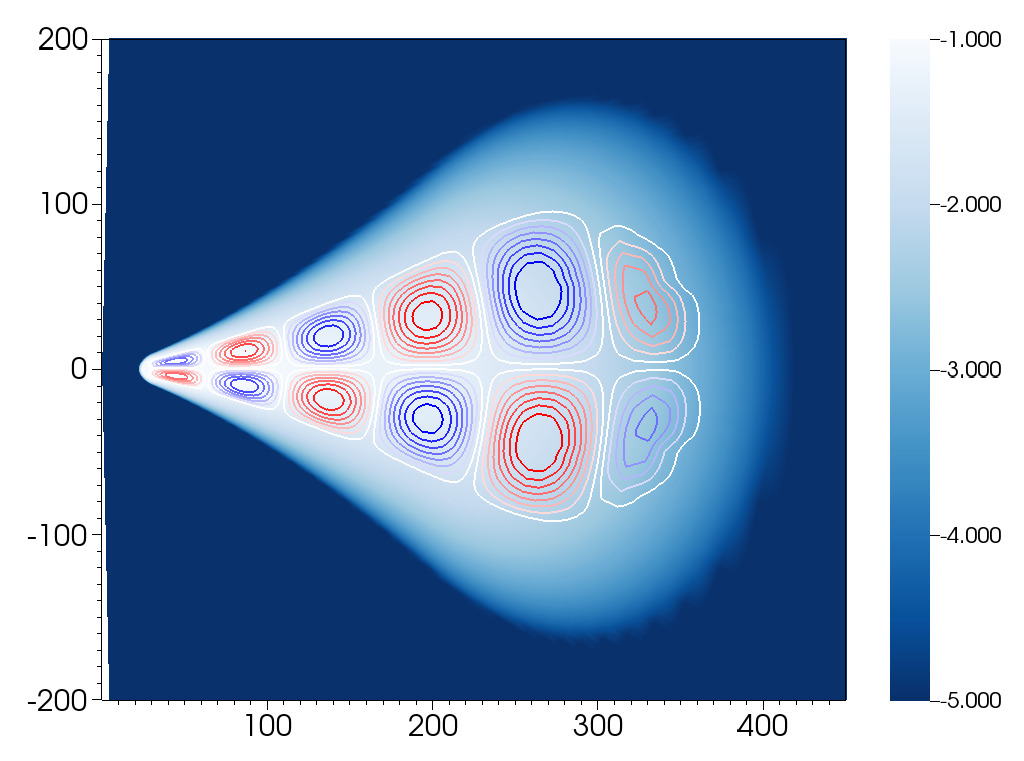}
\includegraphics[width=1.075\columnwidth]{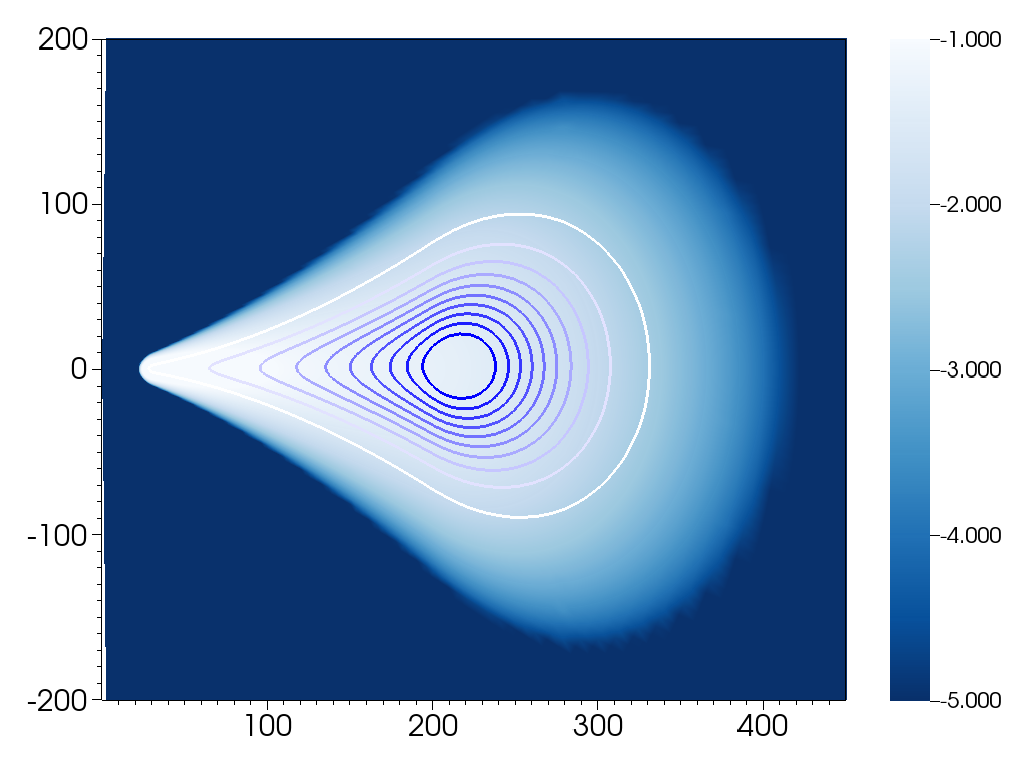}
\caption{Snapshots of logarithm of density for the initial equilibrium
torii for runs \texttt{302a0} and \texttt{302a9mad}. Contours follow the lines of the initial
poloidal magnetic field. Red and blue correspond to clock- and counter-clockwise
loops, respectively.}
\label{f.initial}
\end{figure}

The initial magnetic field threading the torus was purely poloidal
and consisted of either a single or multiple loops. Multiple loops were 
constructed from the vector potential following \cite{penna-limotorus}.
In their notation we adopted 
$r_{\rm start} = 27.5$, $r_{\rm end} = 350$ and $\lambda = 2.5$. To make
the loops flip polarity across the equatorial plane we additionaly multiplied
the vector potential by $\sin (\pi/2-\theta)$. The single loop, on the contrary,
 was set up
according to the vector potential given by,
\be
A_\varphi = {\rm max}\left(\left(\frac {\rho R^2}{6\times 10^{-5}{\rm g/cm^3}\,G^2M^2/c^4}\right)^2-0.02,0\right)
(\sin\theta)^4.
\ee
The initial torus for the two configurations of the
 magnetic field  is shown in Fig.~\ref{f.initial}.

\begin{table*}
\centering
\caption{Model parameters}
\label{t.models}
\begin{tabular}{lcccccccc}
\hline
\hline
Name & $a_*$ & $B$ field & $\cal{K}$ & $\beta_{\rm max}$ & resolution           & grid                                          &  $t_{\rm max}$ & $\langle \dot M \rangle /\Medd$\\
     &       &           &           &                 & ($N_R$ x $N_\theta$)  & ($R_{\rm min}$ / $R_{\rm max}$ / $R_0$ / $H_0$)    &              & \\
\hline
\hline
\multicolumn{8}{l}{super-critical disks, non-rotating BH:}\\
\hline
\texttt{r302a0} & $0.0$ & multi.  & $100$     & $10$            & (264 x 200)          & (1.85 / 5000 / 1.0 / 0.6)              &  128,000  &  559\\ 
\texttt{r3015a0} & $0.0$ & multi.  & $200$     & $10$            & (264 x 200)          & (1.85 / 5000 / 1.0 / 0.6)              &  151,000  &  73.1 \\ 
\texttt{r301a0} & $0.0$ & multi.  & $300$     & $10$            & (264 x 224)          & (1.85 / 5000 / 1.0 / 0.6)              &  111,000  &  24.3 \\ 
\texttt{r300a0} & $0.0$ & multi.  & $500$     & $10$            & (264 x 240)          & (1.85 / 5000 / 1.0 / 0.6)              &  111,000  &  9.6 \\ 
\hline
\multicolumn{8}{l}{super-critical, rotating BH:}\\
\hline
\texttt{r302a9madhb} & $0.9$ & single  & $100$     & $10$            & (240 x 240)          & (1.3 / 5000 / 1.0 / 0.4)              &  90,500  &  3050  \\
\texttt{r302a9mad} & $0.9$ & single  & $100$     & $50$            & (240 x 240)          & (1.3 / 5000 / 1.0 / 0.4)              &  90,500  &  2060  \\
\texttt{r300a9} & $0.9$ & multi.  & $500$     & $10$            & (264 x 240)          & (1.3 / 5000 / 1.0 / 0.6)              &  54,000  &  11.9 \\ 
\hline
\multicolumn{8}{l}{thin, collapsing disks:}\\
\hline
\texttt{r299a0}   & $0.0$ & multi.  & $800$     & $10$            & (264 x 216)          & (1.85 / 5000 / 1.0 / 0.7)              &  155,000  &  2.1$^*$ \\ 
\texttt{r297v3a0} & $0.0$ & multi.  & $1000$     & $10$            & (288 x 300)          & (1.85 / 1000 / 1.0 / 0.85)              &  25,050  &  N/A \\ 
\texttt{r297v2a0} & $0.0$ & multi.  & $1050$     & $10$            & (288 x 312)          & (1.9 / 1000 / 1.25/ 0.9)              &  27,000  &  N/A \\ 
\texttt{r297v1a0} & $0.0$ & multi.  & $1100$     & $10$            & (288 x 300)          & (1.9 / 1000 / 1.25/ 0.9)              &  24,100  &  N/A \\ 
\texttt{r298a9}   & $0.9$ & multi.  & $1000$     & $10$            & (280 x 300)          & (1.3 / 5000 / 1./ 0.85)              &  33,700  &  N/A \\ 
\hline
\hline
\multicolumn{8}{l}{multi. - mutiple magnetic field loops; single - single magnetic loop;}\\
\multicolumn{8}{l}{ $\beta_{\rm max}$ - maximal value
of magnetic $\beta=p_{\rm tot}/p_{\rm mag}$; $t_{\rm max}$ - duration of simulation. }\\
\multicolumn{8}{l}{$^*$ Before $t=70,000$. }
\end{tabular}
\end{table*}

\subsection{Super-critical disks around non-rotating BHs}
\label{s.SANE}

We performed a set of eight simulations with multiple loops of the
initial magnetic field and a non-rotating BH. The intial torus in all simulations
was the same (but for the seed magnetic field), but we varied the entropy parameter ${\cal K}$ responsible for scaling the density.
The four simulations with the highest densities (models
\texttt{r302a0} -- \texttt{r300a0}, see Table~\ref{t.models}) evolved into a 
quasi-stationary states corresponding to accretion rates 
$\dot M/\Medd = 10 \div 560$. The other four did not provide
equilibrium solutions, and collapsed (model \texttt{r299a0} only after
$t\approx 70,000$). They are discussed in Section~\ref{s.unstable}. Below we 
described in detail 
the disks which reached equilibrium. The discussion is extended
to spinning BHs in Section~\ref{s.spinning}.

\subsubsection{General properties}

\begin{figure}
\includegraphics[width=1.0\columnwidth]{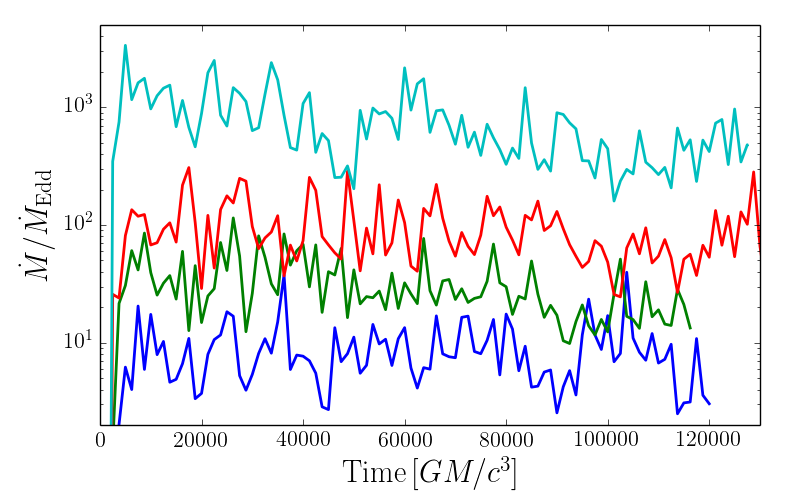}
\caption{Time profiles of the accretion rate through the BH horizon
for models (top to bottom) \texttt{r302a0}, \texttt{r301a0}, \texttt{r3015a0}, and \texttt{r300a0}.}
\label{f.mdotvst.sane}
\end{figure}

Figure~\ref{f.mdotvst.sane} shows the accretion
rate history for the four equilibrium runs. The different accretion rates are mostly due to the 
different starting densities of the initial torus.  Another (less important) factor that affects the 
accretion rate is the resulting radial velocity. 
In the standard disk picture, 
optical depth of the disk (which depends on the
initial torus density) determines how radiatively efficient
it is, and also affects (although slightly in our cases) the radial velocity. 

For all four simulations the accretion rate remains roughly constant
or slightly decreases with time. This proves that the sub-grid
dynamo is effective in keeping the turbulence and accretion alive.
The weak decreasing trend in accretion rate results both
from the depletion of inital gas reservoir, and increasing 
magnetic pressure near the BH.

\begin{figure}
\includegraphics[width=1.0\columnwidth]{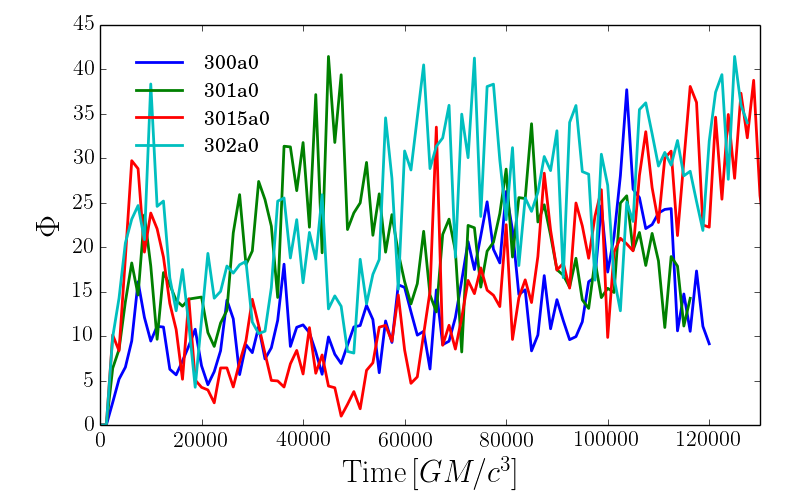}
\caption{Evolution of the magnetic flux parameter $\varphi$ (Eq.~\ref{eq.Phi}) with time.}
\label{f.phivst.sane}
\end{figure}

The history of the magnetic flux accumulated on the BH horizon,
parametrized through,
\be 
\label{eq.Phi}
\varphi = \frac 1{\sqrt{\langle\dot M\rangle}} \frac{4\pi}{2} \int_{0}^\pi
\int_0^{2\pi}\sqrt{-g}\,|B^r|\,{\rm d}\varphi {\rm d}\theta, 
\ee 
where $\langle\dot M\rangle$ is the average accretion rate for
each run, is shown in Figure~\ref{f.phivst.sane}. 
All four runs where initiated with quadrupole magnetic field, so
magnetic field with alternating polarity is expected
to reach BH with time. This results in roughly chaotic
evolution of the magnetic flux parameter. The average
value of $\varphi$ over entire duration of simulations falls
between $\varphi=14.4$ for \texttt{r300a0}, and $\varphi=23.5$ for \texttt{r302a0},
and is significantly below the Magnetically Arrested Disk (MAD)
critical value $\varphi_{\rm MAD}\approx 50$ 
\citep{tchekh10a,tchekh+12}.

\begin{figure}
\includegraphics[width=1.075\columnwidth]{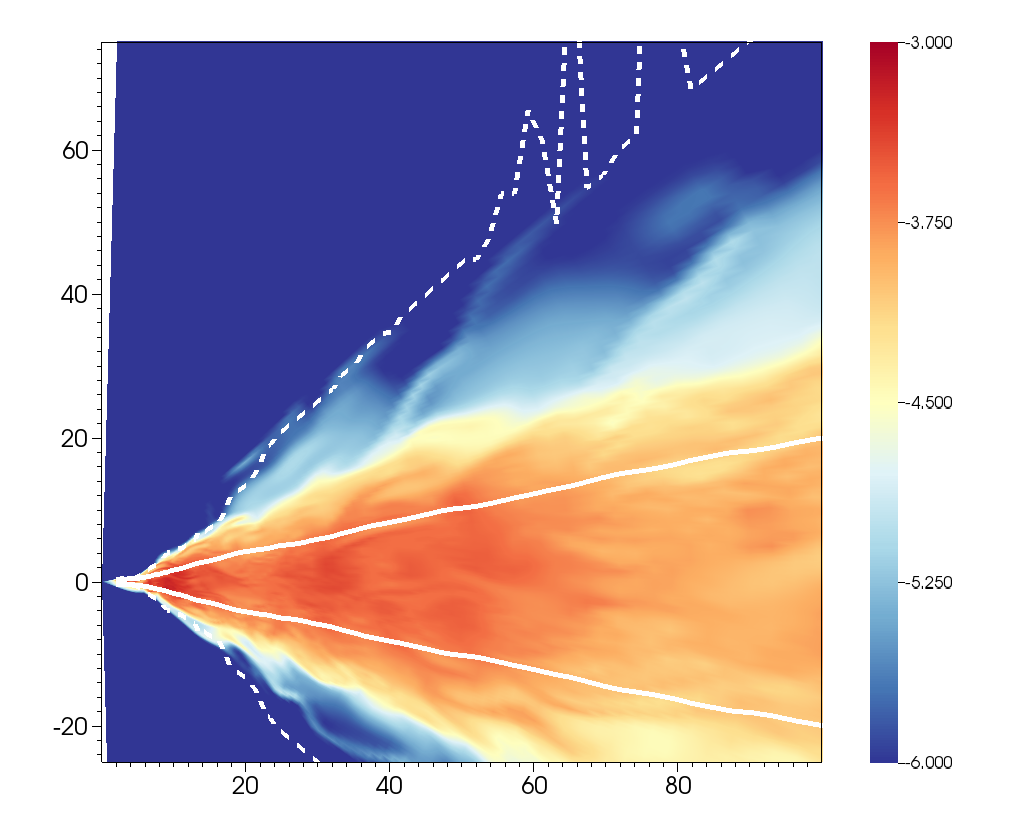}
\includegraphics[width=1.075\columnwidth]{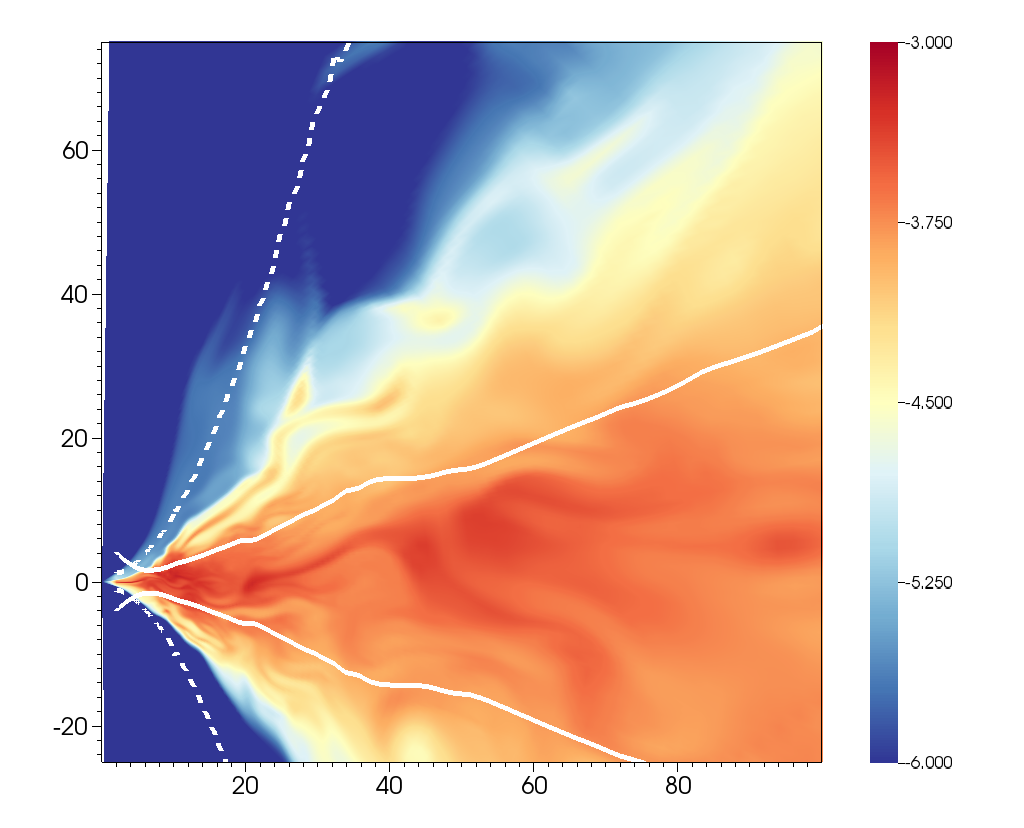}
\includegraphics[width=1.075\columnwidth]{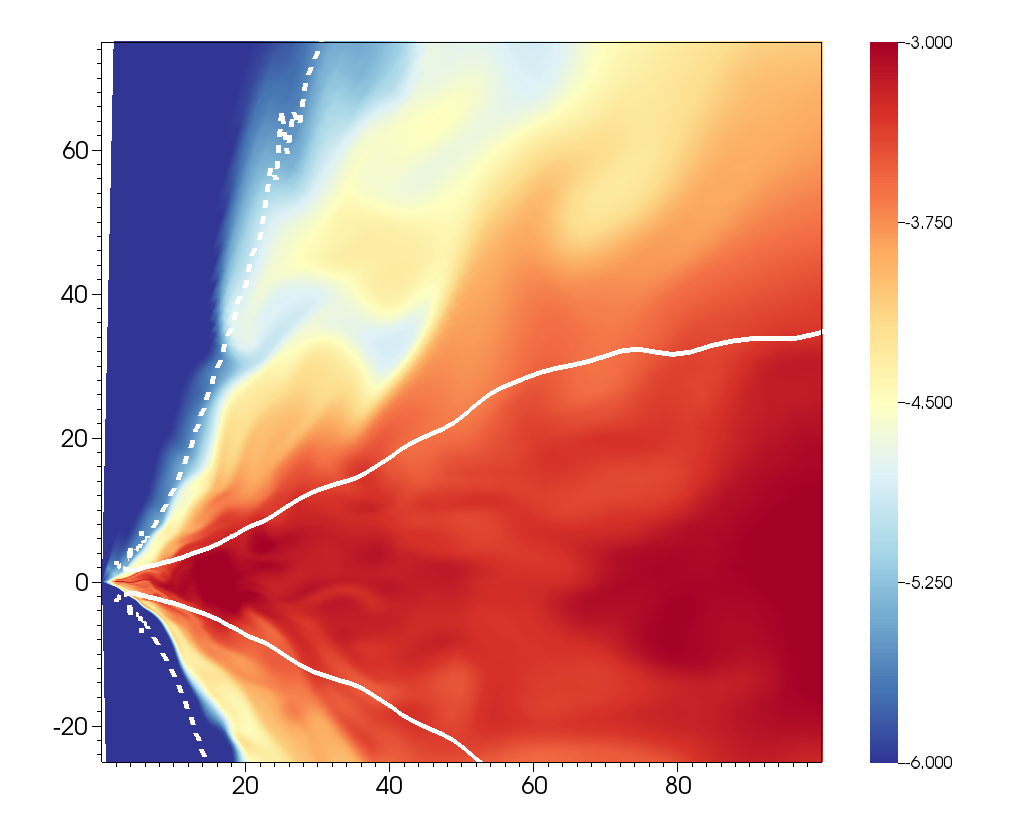}
\caption{Snapshots of logarithm of density for models 
\texttt{r299a0}, \texttt{r300a0}, and \texttt{r301a0} (top to bottom). 
Solid and dashed lines show locations of the density scale height and the photosphere, 
respectively.}
\label{f.triple}
\end{figure}

Figure~\ref{f.triple} shows snapshots from three models,
\texttt{r299a0} (before collapse, see Section~\ref{s.unstable}), \texttt{r300a0}, and
\texttt{r301a0}. Colors show the logarithm of density.
Solid and dashed lines show the scale-height,
\be
\label{eq.scaleheight}
\theta_H = \sqrt{\,\frac{2\pi}\Sigma\int_0^{\pi}\rho |\theta-\pi/2|^2 \sqrt{g_{\theta\theta}}\,{\rm d}\theta},
\ee
where
\be
\label{eq.sigma}
\Sigma=2\pi\int_0^{2\pi}\rho \sqrt{g_{\theta\theta}}\,{\rm d}\theta,
\ee
is the surface density,
and the location of the photosphere\footnote{Plots show the photosphere
surface obtained by integrating the optical depth from the polar axis to $\tau=1$ along
a fixed radius, while in Section~\ref{s.photosphere}, we 
perform the integrals from the outer boundary towards the BH along a constant polar angle.}
respectively. The turbulent structure of each disk
is clear. Increasing accretion rate results both in larger densities, 
and thicker (in terms of the scale-height and the photosphere
location) disks.

\subsubsection{Resolving MRI}

\begin{figure}
\includegraphics[width=1.0\columnwidth]{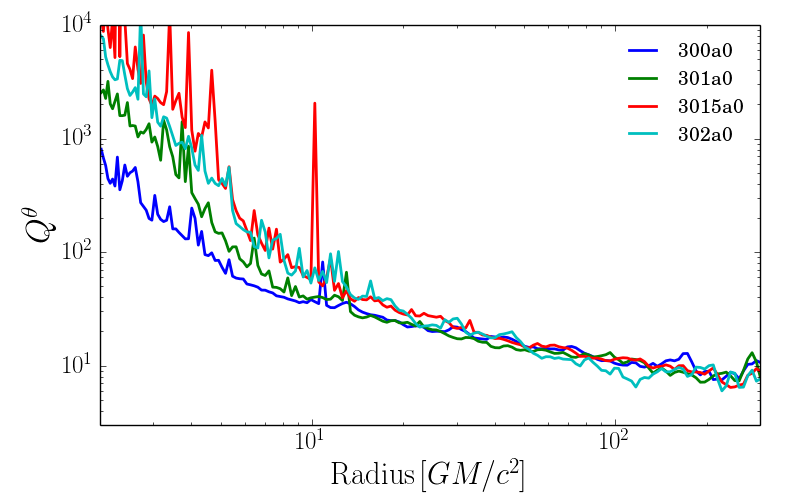}
\caption{Vertically and time averaged MRI resolution parameter $Q^\theta$.}
\label{f.qtheta.sane}
\end{figure}

To verify that the adopted resolution is enough to resolve
the fastest growing mode of the MRI we calculate the vertically averaged parameter
$Q^\theta$ \citep{hawley+11},
\be
\label{eq.qtheta}
Q^{\theta}=\frac{2\pi}\Sigma\int_0^\pi \rho\frac{2\pi}{\Omega \Delta x^\theta}\frac{|B^\theta|}{\sqrt{\rho}}\sqrt{g_{\theta\theta}}\,{\rm d}\theta,
\ee
where $\Delta x^\theta$ is the grid cell size in $\theta$, and $\Omega$ is
the angular velocity. Figure~\ref{f.qtheta.sane} shows profiles of 
the MRI resolution parameter for all the runs with non-spinning BHs 
obtained by averaging values calculated for each snapshot (taken every
$\Delta t=50$). For all radii $R\lesssim 60$, $Q^\theta$ significantly exceeds 10 making
 MRI reasonably resolved \citep{hawley+13}.

\subsubsection{Rest mass fluxes}

\begin{figure}
\includegraphics[width=1.0\columnwidth]{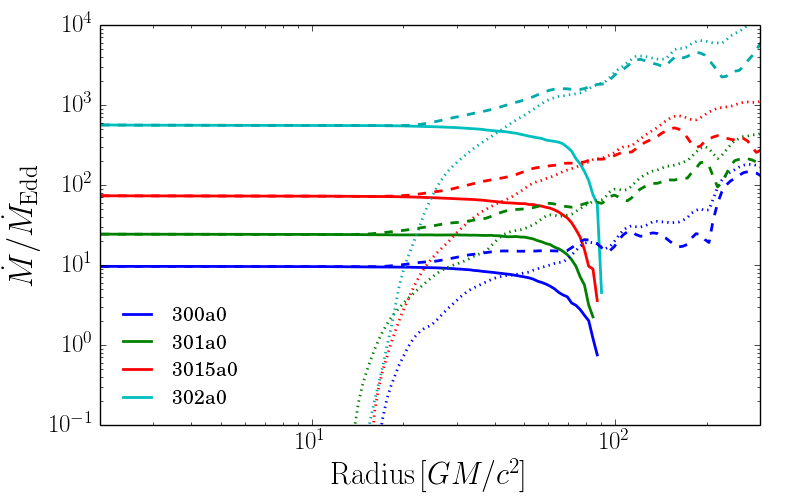}
\caption{Radial profiles of the rest mass flux (accretion rate). Solid
lines show the total flux, while dashed and dotted lines denote the
fluxes of inflowing, and outflowing gas only, respectively.}
\label{f.mdotvsr.sane}
\end{figure}

Figure~\ref{f.mdotvsr.sane} shows the averaged profiles of the
total (solid), inflowing (dashed), and outflowing (dotted lines)
accretion rate. The total accretion rate for all four runs
is roughly constant between the BH horizon and radius $R_{\rm eq}\approx 60$,
which limits the range of the inflow-outflow equilibrium. 
The inflowing accretion rate departs from the net accretion rate
 outside radius $R_{\rm out}\approx 20$, where gas starts to 
flow out in the wind region (dotted lines). The wind
mass loss rates equal the accretion rates near radius
$R\approx 40$ for all four runs. The radial slope of 
the 
inflowing accretion rate outside $R_{\rm eq}$ is close to $\dot M_{\rm in}\propto R^1$, but
is rather poorly constrained because of
the limited range of the equilibrium solution.

\cite{narayan+12} and \cite{sadowski+outflows} discussed properties
of the rest mass flux in three-dimensional simulations of
radiatively inefficient, optically thin accretion disks.
For their simulations with a non-rotating BH, no significant
mass outflow was present inside $R\approx 50$ (compared to 
our $R\approx 20$). Similar conclusions may be drawn from comparing
simulations with spinning BHs (Fig.~\ref{f.mdotvsraN}). These
facts suggest that optically thick, radiation pressure dominated
disks produce winds more effectively than optically thin disks.

\subsubsection{Energy fluxes}
\label{s.energyfluxes}

\begin{figure}
\includegraphics[width=1.0\columnwidth]{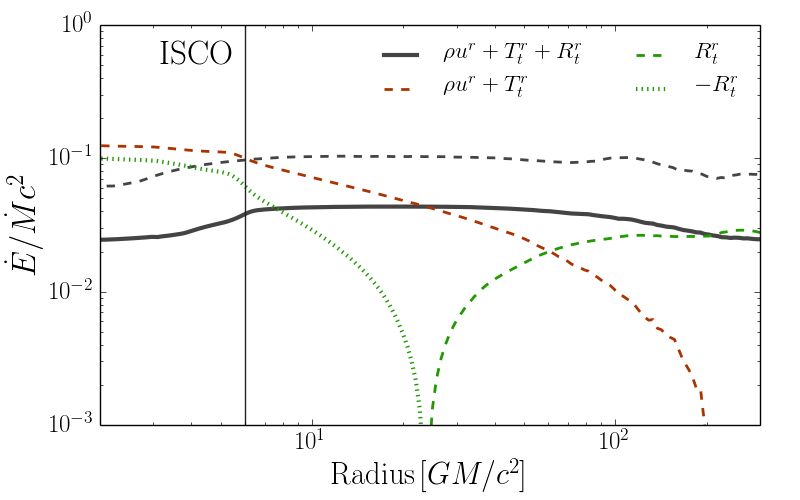}
\caption{Radial profiles of the energy fluxes for simulation
  \texttt{r300a0}. The black solid line shows the flux of the total
energy outflowing to infinity. Red and green lines show the MHD and radiative
components, respectively. Dashed and dotted sections correspond to 
outflowing and inflowing energy flux, respectively. Dashed black line shows the
total energy flux for run \texttt{r300a9}.}
\label{f.enflux.sane}
\end{figure}

Figure~\ref{f.enflux.sane} shows the averaged profiles 
of the energy fluxes normalized by the rest mass accretion 
rate at the horizon, and integrated over the full solid angle,
for a representative run \texttt{r300a0}.
The solid black line shows the total flux of energy available 
at infinity obtained by summing up the MHD and radiative fluxes
($T^r_t$ and $R^r_t$, respectively), and substracting the
rest mass energy flux ($\rho u^r$), normalized by the rest
mass energy flux through the BH horizon. The profile of the total energy flux is flat between 
the ISCO and the convergence radius $R\approx 60$, and corresponds 
roughly to efficiency of $4\%$ --- this fraction of $\dot M c^2$ is
extracted by accretion and ejected as a sum of radiative flux,
magnetic Poynting flux, and kinetic and thermal energies. This 
number is only slightly lower than the efficiency implied by the standard model 
of a radiatively efficient thin disk ($\eta_0=5.7\%$) and is roughly
constant for all the $a_*=0.0$ runs (see Table~\ref{t.luminosities}).

The total flux of energy can be decomposed into the MHD 
(magnetic, kinetic, and thermal components), and radiative 
(energy carried by photons) fluxes. At large radii the radiative
flux dominates over the MHD flux because it consists of 
accumulated contributions of all the energy
liberated by viscosity in the inner region.
The radiative flux approaches $\dot E_{\rm rad}=0.03\dot Mc^2$, what
for accretion rate $\dot M\approx 10$, and radiative efficiency of a
thin disk $\eta_0\approx 0.06$ (Eq.~\ref{e.medd}), gives total radiative
luminosity $L\approx 0.03 \times 10 / 0.06\,L_{\rm Edd}=5 \,L_{\rm Edd}$.
The net radiative flux of energy changes sign inside $R\approx 20$,
reflecting
the fact that in this regions photons are effectively trapped and most
of them are advected onto the BH. The departure of the total net
energy flux from otherwise constant value $\sim 4\%$ inside ISCO is caused by
rapidly increasing radial velocity of the flow in the plunging region,
which increases the numerical diffusivity of the  scheme and
introduces
extra viscous contribution to the energy flux not reflected in the
quantities
that we plot.

More detailed discussion of disk luminosities is given in
Section~\ref{s.luminosities}.

\subsubsection{Magnetic field properties}

\begin{figure}
\includegraphics[width=1.0\columnwidth]{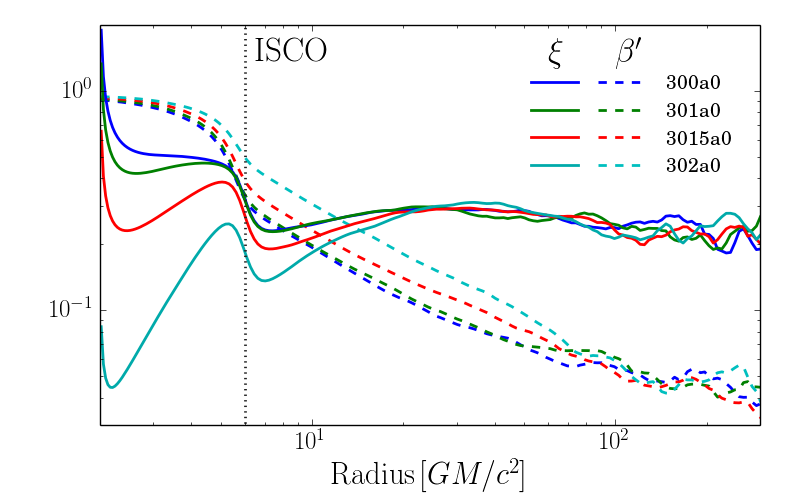}
\caption{Average parameters of the magnetic field. The dashed lines show
the magnetic to total pressure ratio, and the solid lines show
the magnetic field angle (Eq.~\ref{eq.bangle}).}
\label{f.betatheta.sane}
\end{figure}

The poloidal magnetic field is prevented from decaying
by the sub-grid dynamo described in Appendix~\ref{ap.dynamo}. In all the models, we 
set
the target magnetic field angle $\xi_{\rm dyn}$ (Eq.~\ref{eq.bangle})
to $0.25$, and the magnetic to total pressure ratio
$\beta'_{\rm damp}$ to $0.1$. Figure~\ref{f.betatheta.sane}
shows the radial profiles of the averages (vertical
with density weight, and over time) of these two quantities.
The dashed lines show the magnetic field angle $\xi_{\rm dyn}$ which was
expected to settle down near $0.25$. Indeed, for all runs
and $R\gtrsim10$ the mean magnetic field angle is close to that
value. It departs from $\xi\approx0.25$ only in the innermost region, where the 
magnetic field is determined by the rapidly accelerating and sheared gas.
Solid lines in Figure~\ref{f.betatheta.sane} show the average ratio
of magnetic to total pressure. Whenever it exceeds the prescribed critical 
value $\beta'_{\rm damp}=0.1$, the toroidal magnetic field is damped.
Because of this asymmetry (no action is taken when $\beta'_{\rm damp}<0.1$) the 
resulting average $\beta'$ is expected to be 
lower than $0.1$. It is the case for the region outside $R\gtrsim 60$,
where the pressure ratio for all the simulations is around $\beta'=0.05$.
Inside the inflow equilibrium region, the magnetic field component 
contributes more to the total pressure with decreasing radius. The 
typical value of that pressure ratio at $R=10$ is $0.2-0.3$. It exceeds
the critical value of $0.1$ because the damping of the toroidal
component of magnetic field becomes less and less effective when
the radial velocity (see Fig.~\ref{f.vrvsr.sane}) is large. The shear,
which stretches the poloidal component of the magnetic field and 
amplifies the toroidal one, is unaffected by radial velocity. The
sub-grid damping, however, acts on the orbital timescale (Eq.~\ref{eq.dynamodamp}),
and gas moving with larger velocity is affected by damping over a shorter time,
making it less effective. The resulting profile of the magnetic pressure is, however,
in qualitative agreement with 3D simulations of ADAFs \citep[e.g.,][]{narayan+12},
which also show increased contribution of magnetic pressure towards the BH.

\begin{figure}
\includegraphics[width=1.0\columnwidth]{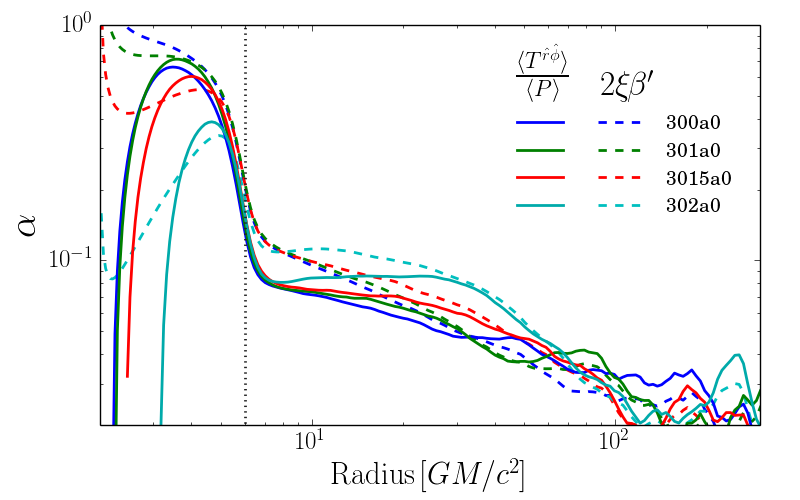}
\caption{
Profiles of the viscosity parameter $\alpha$. Dashed lines show the
estimate based on the product of the magnetic field angle and pressure ratio. Solid
lines show the estimate of $\alpha$ calculated consistently.}
\label{f.alphavsr.sane}
\end{figure}

Figure~\ref{f.alphavsr.sane} shows the radial profiles of the
turbulent viscosity parameter $\alpha$ \citep{ss73}. The dashed
lines show the crude estimate, based on the product of the 
magnetic field angle and pressure ratio,
\be
\alpha=\frac{\langle T^{\hat r \hat \varphi}\rangle}{\langle p \rangle}\approx
\frac{\langle b^{ r} b^{ \varphi}\rangle}{\langle p \rangle}=
\frac{\langle b^{ r} b^{ \varphi}\rangle}{\langle b^2\rangle}\frac{\langle b^2\rangle}{\langle p \rangle}
=2\langle\xi\rangle \langle\beta'\rangle,
\ee
while the solid lines show the viscosity parameter $\alpha$ estimated
in a more consistent way directly through, 
\be
\label{eq.alpha}
\alpha=\frac{\langle T^{\hat r \hat \varphi}\rangle}{\langle p \rangle},
\ee
where $T^{\hat r \hat \varphi}$ is the ortonormal $(r,\varphi)$ component of the
gas stress-energy tensor in the mean comoving frame, and $p$ is the total pressure. Following 
\cite{penna+alpha} we choose the azimuthal component of the gas velocity
as the only non-zero component of the comoving frame at given location.
The averages are taken as before, but this time only within one
scale height of the disk to avoid strongly magnetized corona.

The two ways of estimating the viscosity paramer show good agreement. The $\alpha$ estimated with
 Eq.~\ref{eq.alpha} has relatively flat profile between ISCO and $R\approx 60$,
and falls in the range $\alpha=0.06 \div 0.1$. It rapidly grows inside ISCO,
reflecting the increase of the magnetic pressure, and falls down again when approaching the
horizon. The whole profile is also in good qualitative agreement with 
the characteristics of $\alpha$ observed in 3D simulations, although the very
 rapid grow of $\alpha$ at ISCO is rather unphysical \citep{penna+alpha}.

\subsubsection{Surface density, optical depth and velocities}

\begin{figure}
\includegraphics[width=1.0\columnwidth]{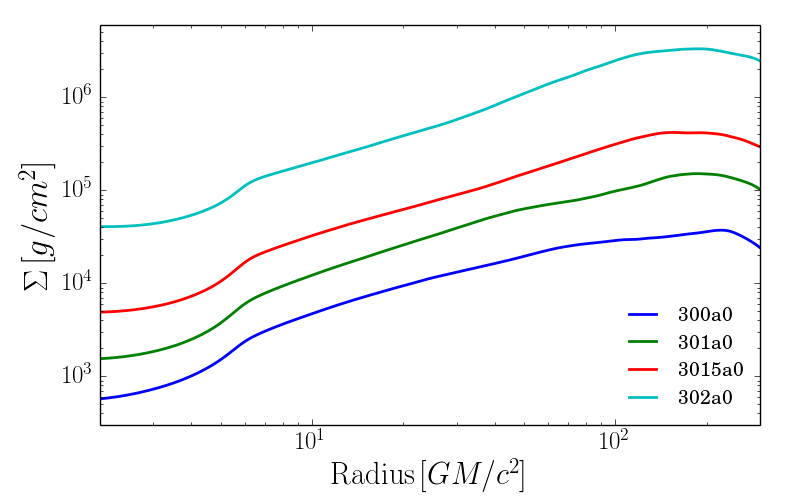}
\caption{Radial profiles of the surface density.}
\label{f.sigmavsr.sane}
\end{figure}

\begin{figure}
\includegraphics[width=1.0\columnwidth]{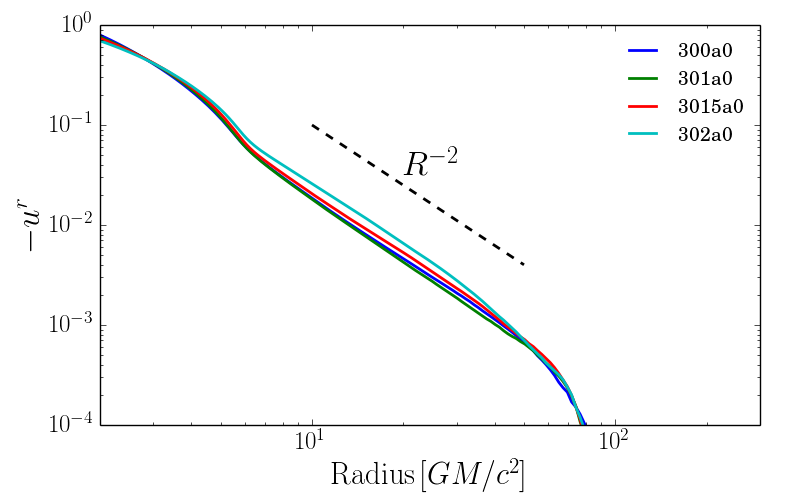}
\caption{Radial profiles of the average radial velocity $<\rho u^r>/<\rho>$.}
\label{f.vrvsr.sane}
\end{figure}

Figures~\ref{f.sigmavsr.sane} and \ref{f.vrvsr.sane} show profiles of
the surface density and radial velocity, respectively.
Surface densities at given radius increase with accretion 
rates suggesting that the disk solutions correspond to the
top, advection-dominated branch of slim disks on so-called S-curve diagrams
\citep{abra88}. Outside ISCO, surface density increases roughly
proportionally to radius, what is also in good agreement
with slim disk models \citep[see e.g.,][]{sadowski.phd}.

\begin{figure}
\includegraphics[width=1.0\columnwidth]{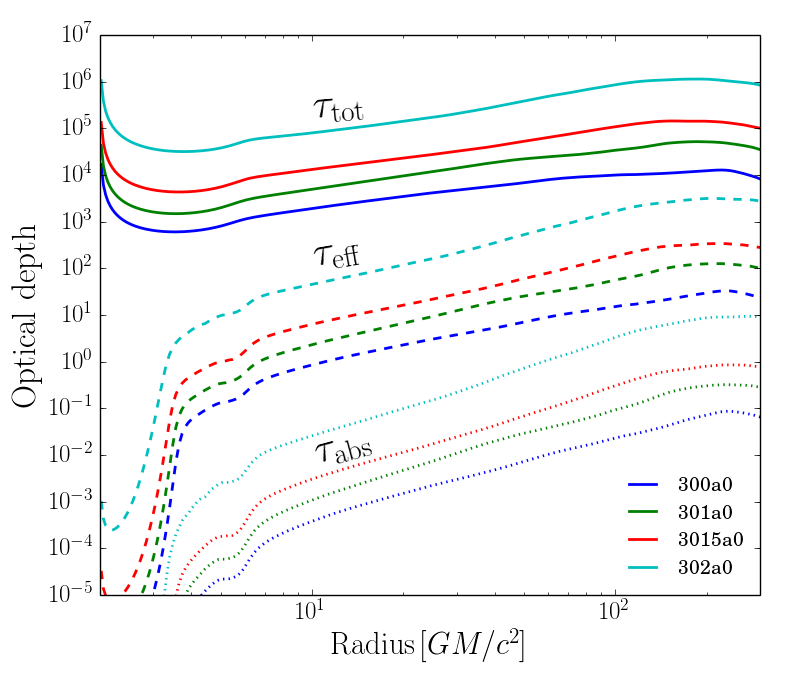}
\caption{Radial profiles of the optical depth in
simulations \texttt{r300a0} - \texttt{r302a0}. Solid,
dashed and dotted lines correspond to the total, 
effective, and absorption optical depths, respectively.}
\label{f.tauvsr.sane}
\end{figure}

Figure~\ref{f.tauvsr.sane} presents radial profiles of the optical
depth. The total optical depth, defined as, \be \tau_{\rm
  tot}=\int_0^{\pi}\rho(\kappa_{\rm es}+\kappa_{\rm a})u^t
\sqrt{g_{\theta\theta}}\,{\rm d}\theta, \ee is shown with solid
lines. Because scatterings strongly dominate over absorptions, the
profiles of the total optical depth roughly correspond to the surface
density profiles rescaled by $\kappa_{\rm es}=0.34\,\rm cm^2/g$. Only
close to the horizon the extra Lorentz factor $u^t$ affects the
scalings. For all the four simulations discussed here
(\texttt{r300a0}\,--\,\texttt{r302a0}) the total optical depth exceeds
$\tau_{\rm tot}=1000$, making them very optically thick.  They are,
however, at the same time optically thin with respect to absorptions,
because of high temperatures in the disk body. In this context, an
important measure is the effective optical depth, which determines if
a photon is absorbed along its scattering-affected path.  We
approximate the effective optical depth by calculating, 
\be \tau_{\rm
  eff}=\sqrt{(\tau_{\rm tot}+\tau_{\rm abs})\tau_{\rm tot}}, 
\ee where
$\tau_{\rm abs}$ is the optical depth for absorption. The resulting
profiles are plotted in Figure~\ref{f.tauvsr.sane} with dashed
lines. For the three runs with highest accretion rates, $\tau_{\rm
  eff}$ exceeds $1$ everywhere outside ISCO. Simulation
\texttt{r300a0}, however, is effectivelly optically thin ($\tau_{\rm
  eff}\lesssim1$) inside $R\lesssim 10$. This fact may have
significant impact on the emitted spectrum, but quantifying it will
require detailed radiative transfer calculations.

Figure~\ref{f.vrvsr.sane}
shows profiles of the average radial velocity. All the 
simulations show very similar radial velocities
which decrease with radius as $R^{-2}$, consistent
with the slope obtained from simulations of non-radiative,
optically thin disks \citep{narayan+12}. However, there is a 
slight trend of decreasing radial velocity with accretion rate,
which implies that the solutions are not completely in the 
ADAF regime where no dependence on accretion rate is expected.

\begin{figure}
\includegraphics[width=1.0\columnwidth]{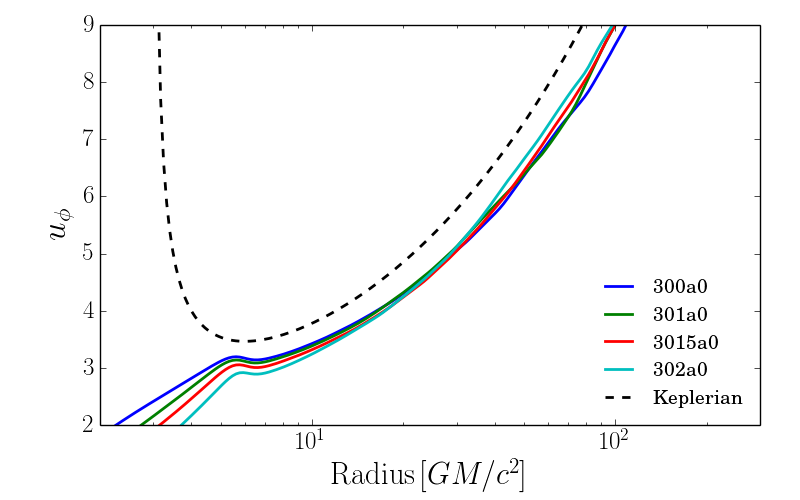}
\caption{Radial profiles of the specific angular momentum $u_\varphi$.}
\label{f.uphivsr.sane}
\end{figure}

Figure \ref{f.uphivsr.sane} shows radial 
profiles of the specific angular momentum. Outside $R\approx10$,
all the simulated disks show angular momentum 
corresponding to roughly $87\%$ of the Keplerian 
angular momentum. Inside $R\approx10$ the profiles 
differ a bit, with the one corresponding to the lowest
accretion rate (\texttt{r300a0}) being the closest
to the Keplerian profile. Inside the ISCO, the angular momentum
drops down for all the runs. This reflects the fact that 
viscosity in the plunging region is no longer regulated
by the dynamo sub-grid model, and the $\alpha$-viscosity parameter
rapidly increases towards BH (Fig.~\ref{f.alphavsr.sane}).

\subsubsection{Photosphere}
\label{s.photosphere}

Radiation generated by viscous dissipation is absorbed, re-emitted and
scattered inside the bulk of the disk. However once a photon crosses the
photosphere, it can escape to an observer located at infinity (the escape path follows a null geodesic).

In this subsection we are primarily interested in the location of the
photosphere in the funnel region close to the axis. Here, gas flows
out at considerable speed. Hence, the numerical solution achieves
steady state out to quite a large radius.  In many cases, inflow equilibrium is established out to the
edge of the box (see the values of $R_f$ in Table~\ref{t.luminosities}). Hence, we can
follow the location of the photosphere to fairly large radii. In
contrast, the discussion of disk properties close to the equatorial
plane in previous subsections was limited to a much smaller range of
radius $R \lesssim R_{\rm eq} \approx 60$.

We estimate the radius $R_{\rm photo}$ of the photosphere in the
funnel as a function of polar angle $\theta$ by integrating the
optical depth from the outer boundary of the simulation box down
towards the BH along fixed $\theta$. In computing the optical depth,
we need to allow for relativistic effects such as GR and motion of the
gas. We make use of the fact that the opacity in any given frame
scales as $1/\omega$, where $\omega$ is the frequency of the radiation
as measured in that frame. We are interested in the opacity in the lab
frame for an observer at infinity, whereas the opacities $\kappa_{\rm
  es}$ and $\kappa_{\rm a}$ correspond to the fluid frame. To carry
out the transformation from one frame to the other, we model a
radially outgoing light ray by an effective wave-vector
$k^\mu=(1,1,0,0)$ and note that the frequency of the radiation as
measured in a frame moving with four-velocity $u^\mu$ is equal to
$-k^\mu u_\mu$. The optical depth at a given $\theta$ from any radius
radius $R$ inside the simulation box to the outer radius of the box
$R_{\rm max}=5000$ is then given by \be
\label{e.photosphere}
\tau_1(R)=-\int_{R}^{R_{\rm max}}\rho(\kappa_{\rm a}+\kappa_{\rm
  es})(u_t+u_r)\sqrt{g_{rr}}\,{\rm d} R', \ee where we use
time-averaged quantities symmetrized with respect to the equatorial
plane. Because $\tau_1(R)$ may significantly underestimate the opacity
to infinity when $R$ is close to $R_{\rm max}$, we define a second
measure of the opacity \be
\label{e.photosphere2}
\tau_2(R)=-[\rho(\kappa_{\rm a}+\kappa_{\rm
    es})(u_t+u_r)\sqrt{g_{rr}}\,R]_R. \ee We then estimate the
photospheric radius $R_{\rm photo}(\theta)$ at the given $\theta$ by
the condition 
\be {\rm max}\left(\tau_1(R_{\rm photo}),\tau_2(R_{\rm photo})\right) = 2/3.  
\ee 
Any value of $R_{\rm photo}$ which is larger than the maximum
radius $R_f$ of steady state in the funnel is discarded. Here, $R_f$
is defined by the condition $R_f/v_r(R_f)=t/2$ where $t$ is the
duration of simulation data from which time-averaged quantities are
obtained for calculating $\tau_1$, $\tau_2$ (typically $t\sim t_{\rm
  max}/2$, where $t_{\rm max}$ is given in Table~\ref{t.models}).

\begin{figure}
\includegraphics[width=1.0\columnwidth]{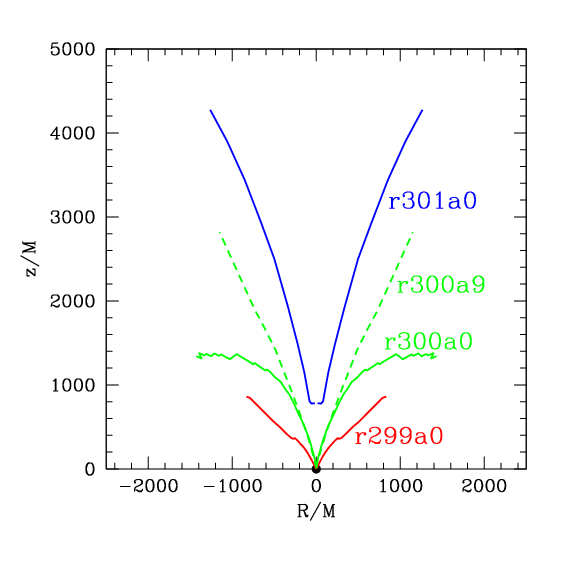}\vspace{-1cm}
\caption{Profiles of the photosphere for disks with non-rotating BHs
  (solid lines), and simulation \texttt{r300a9} ($a_*=0.9$, dashed green line).}
\label{f.photospheres}
\end{figure}

Figure~\ref{f.photospheres} shows the location of the photosphere for
several of our simulated disk models. The red line corresponds to run
\texttt{r299a0} which accreted at $\sim 2\Medd$. Its photosphere is
relatively close to the equatorial plane, but significantly above the
disk density scale height ($\sim0.2$).  The higher the accretion rate,
the closer to the polar axis the photosphere is, and the more
well-defined the optically thin funnel is. For run \texttt{r301a0},
the base of the funnel is at $z=1000$, quite far from the BH, while for
runs \texttt{r3015a0} and \texttt{r302a0}, the whole domain out to
$R_{\rm max}=5000$ is optically thick at all $\theta$. Hence these two
models have no optically thin funnel within the simulation box.  Note
that the optical depth in the funnel is dominated by electron
scattering. Only deep in the optically thick interior of the disk is
absorption opacity important.

The dashed green line in Figure~\ref{f.photospheres} shows the
photosphere for run \texttt{r300a9} which has a comparable accretion
rate as run \texttt{r300a0} but corresponds to a spinning BH with
$a_*=0.9$.  Here the photosphere is located significantly
closer to the polar axis. This reflects the fact that disks with
spinning BHs eject stronger winds (see next Section) and thus tend 
more easily to fill the funnel region with optically thick outflowing
gas.

\subsection{Super-critical disks around rotating BHs}
\label{s.spinning}

In addition to the four runs with non-rotating BHs discussed in the 
previous section, we performed a set of three simulations with
BH spin parameter $a_*=0.9$. One of the most exciting aspects 
of accretion disks around rotating BHs is the possibility of ejecting
jets through the Blandford-Znajek mechanism \citep{bz,penna+bz,lasota+bz}. It has been shown 
both in 2D and 3D that super-critical, optically thick disks can
produce
jets with very similar efficiencies to radiatively inefficient,
optically thin disks \citep{sadowski+koral,mckinney+harmrad}. An open
question is how effective is jet production for accretion disks
accreting at accretion rates comparable to and below the Eddington limit.
In this paper we do not specifically address this question, but study
general properties of outflows from axisymmetric accretion disks
supported
by sub-grid dynamo.

\begin{figure}
\includegraphics[width=1.0\columnwidth]{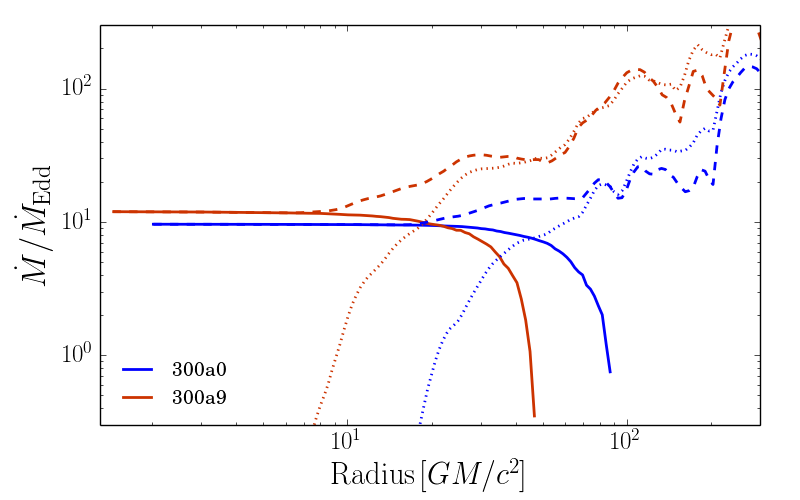}
\caption{Similar to Fig.~\ref{f.mdotvsr.sane} but for two, otherwise
  identical, runs with different BH spin. Blue and red lines
  correspond to runs \texttt{r300a0} and \texttt{r300a9},
  respectively.}
\label{f.mdotvsraN}
\end{figure}

We start with comparing two simulations initiated with the same torus,
and the same intitial magnetic field, but evolved
with different values of BH spin (runs \texttt{r300a0} and
\texttt{r300a9}).
Figure~\ref{f.mdotvsraN} compares the net, inflow and outflow
accretion
rates. The solid lines denote the net rest mass flux
and show that both simulations accrete at roughly the same
level\footnote{In terms of the Eddington accretion rate units defined
  in Eq.~\ref{e.medd}. The 
absolute accretion rate is larger for \texttt{r300a0} by the ratio
of thin disk efficiencies $\eta_0(a_*=0.9) / \eta_0(a_*=0) = 2.74$.}. However, there is
significant difference in the inflowing and outflowing fluxes. While
for the $a_*=0$ run, the gas is lost on the way towards the BH
only down to $R\approx 20$, in case of the spinning BH outflows occur
down to $R\approx 10$. This is consistent with results obtained 
earlier both for optically thin ADAFs, and super-critical optically
thick
disks \citep{sadowski+outflows,sadowski+koral2}, and results
from BH spin impact on the spacetime, which also implies 
the location of ISCO much closer to the horizon for $a_*=0.9$.

\begin{figure}
\includegraphics[width=1.0\columnwidth]{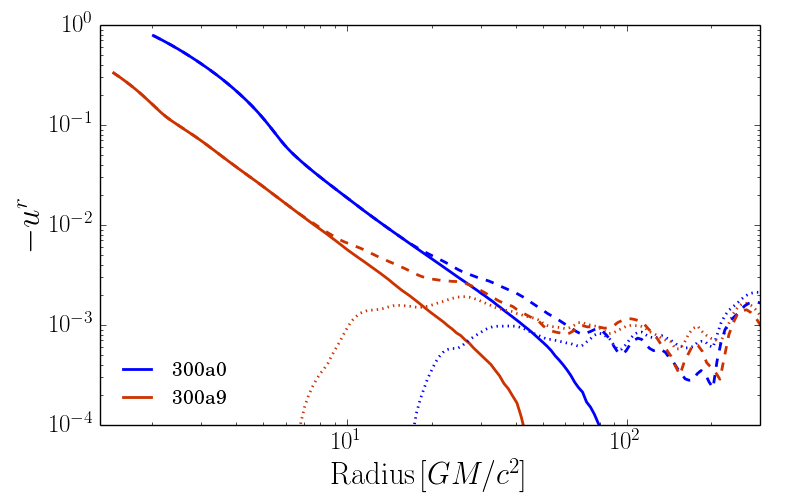}
\caption{Net (solid), inflowing (dashed) and outflowing (dotted lines) radial velocities.
Blue and red lines correspond to runs \texttt{r300a0} and \texttt{r300a9}, respectively.}
\label{f.vrvsraN}
\end{figure}

Figure~\ref{f.vrvsraN} compares radial velocities in the two runs.
The net inward radial velocity is significantly smaller for the run
with spinning BH. This results mostly from the fact that although the
dynamo parameters were set in exactly the same way, the resulting
$\alpha$ viscosity parameter for the $a_*=0.9$ run is roughly $30\%$
lower than for the non-rotating BH case (the inflow velocity is
expected to be proportional to $\alpha$). Additionaly, the magnetic
flux accumulated at the horizon (although relatively small, as will be
discussed in a moment), contributes significantly to the outflowing
energy flux (compare solid and dashed black lines in
Figure~\ref{f.enflux.sane}) because of the Blandford-Znajek process,
slowing the inflowing gas down.

\subsubsection{Magnetic field and accretion efficiency}
\label{s.accreff}

The efficiency of the energy extraction from a rotating BH through the
Blandford-Znajek process depends mostly on the BH spin and the
magnetic flux accumulated on the horizon. It is often characterized by
the magnetic flux parameter $\varphi$ (Eq.~\ref{eq.Phi}) shown in
Figure~\ref{f.vrvsraN} for runs with $a_*=0.9$.  The first two panels
correspond to simulations initiated with a single poloidal loop of
magnetic field. As a result, the magnetic flux accumulates at the
horizon more efficiently than for simulations initiated with a
quadrupolar magnetic field (bottommost panel and
Figure~\ref{f.phivst.sane}).  From the two simulations initiated with
a single loop, magnetic flux accumulates much faster for run
\texttt{r302a9madhb} which had stronger initial seed magnetic
field. We stopped that simulation once the $\varphi$ parameter
exceeded $\varphi=50$, which corresponds to the Magnetically Arrested
Disk (MAD) limit which cannot be simulated properly in
axisymmetry\footnote{In 3D gas can make its way through the accumulated
magnetic field by breaking axisymmetry via an interchange instability,
\cite[e.g.,][]{linarayan-04}, but this is not possible in 2D. Hence 2D
simulations cannot be trusted once they reach the MAD limit.)}.

\begin{figure}
\includegraphics[width=1.0\columnwidth]{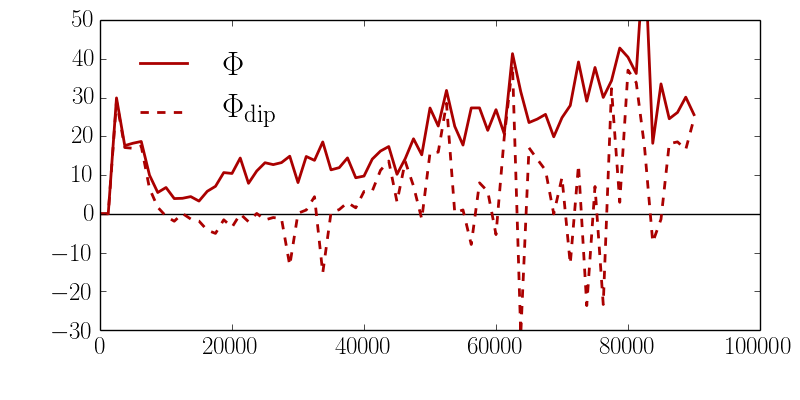}\vspace{-.45cm}
\includegraphics[width=1.0\columnwidth]{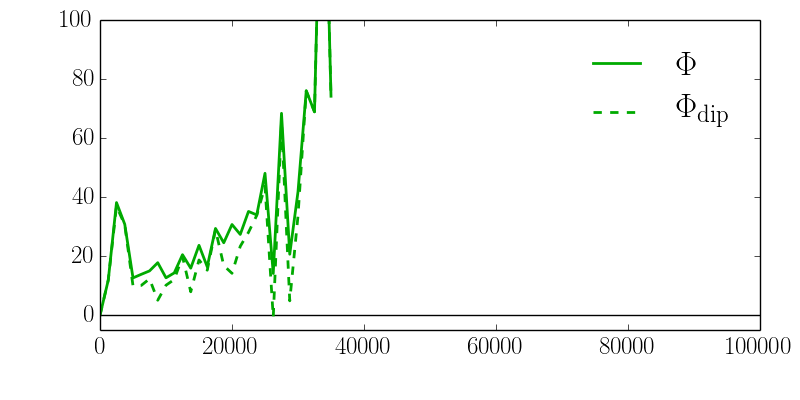}\vspace{-.45cm}
\includegraphics[width=1.0\columnwidth]{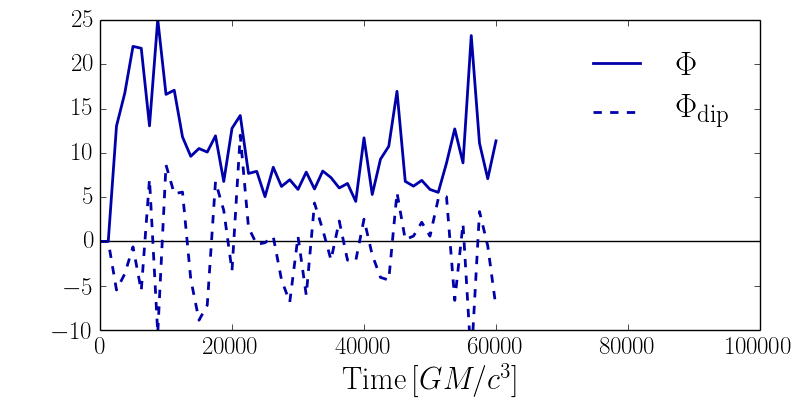}
\caption{Total ($\varphi$, Eq.~\ref{eq.Phi}) and dipolar ($\varphi_{\rm dip}$, Eq.~\ref{eq.Phiquad}) magnetic flux parameter calculated at the BH
horizon for runs (top to bottom) \texttt{r302a9mad},
\texttt{r302a9madhb}, and \texttt{r300a9}.}
\label{f.phivst}
\end{figure}

The definition of the parameter $\varphi$ does not discriminate between 
the topology of the magnetic field, as it integrates the absolute
value
of the radial component of the magnetic field over the horizon. To 
get insight into the topology of the magnetic field crossing the
horizon we define parameter $\varphi_{\rm dip}$,
\be 
\label{eq.Phiquad}
\varphi_{\rm dip} = \frac 1{\sqrt{\langle\dot M\rangle}} \frac{4\pi}{2} \int_{0}^\pi
\int_0^{2\pi}\sqrt{-g}\,B^r\,{\rm sign}(\theta-\pi/2)\,{\rm d}\varphi {\rm d}\theta, 
\ee 
For a perfectly dipolar magnetic field at the horizon one has
$|\varphi_{\rm dip}|=\varphi$. For a quadrupolar (or any even-polar)
magnetic field, $\varphi_{\rm dip}=0$. This parameter is plotted with
dashed
lines in Figure~\ref{f.phivst}. For simulations \texttt{r302a9mad} 
(top panel, initiated with a single loop of weak magnetic field)
and \texttt{r300a9} (bottom panel, initiated with multiple loops),
$\varphi_{\rm dip}$ oscillates around zero showing that the magnetic
field
at the horizon is close to quadrupolar, and that it often flips
orientation.
This is because the dynamo generates turbulent magnetic
field and, in case of simulation \texttt{r302a9mad}, quickly 
overwhelmes the initial dipolar field. For run \texttt{r302a9madhb} 
(middle panel), on the contrary, $\varphi_{\rm dip}$ stays all the time
close to $\varphi$, and does not change sign, because the magnetic
field at the horizon is purely dipolar. This results from stronger
initial magnetic field which is not so easily overwritten by dynamo-generated
turbulent field.

\begin{figure}
\includegraphics[width=1.0\columnwidth]{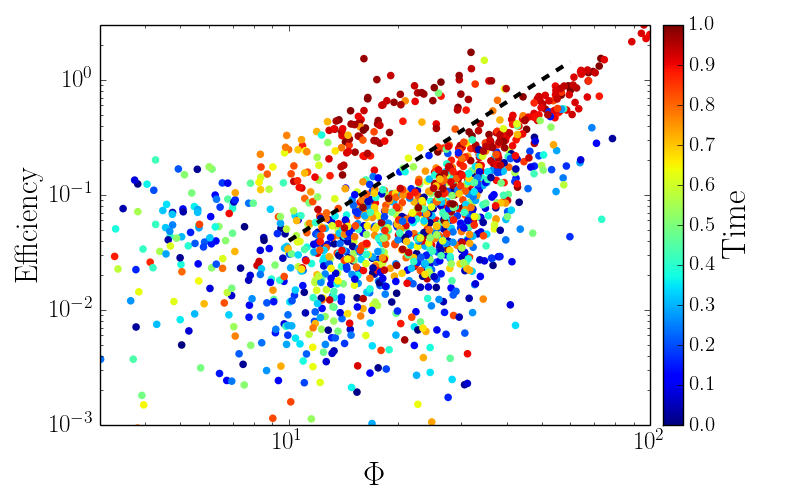}\vspace{-.45cm}
\includegraphics[width=1.0\columnwidth]{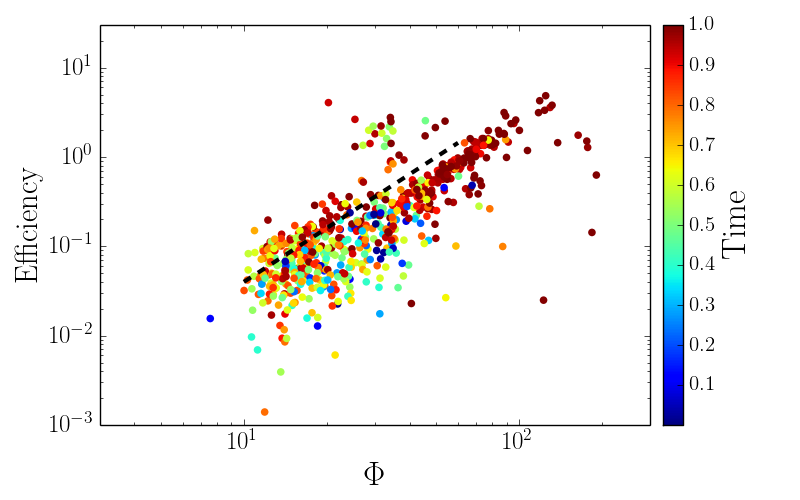}\vspace{-.45cm}
\includegraphics[width=1.0\columnwidth]{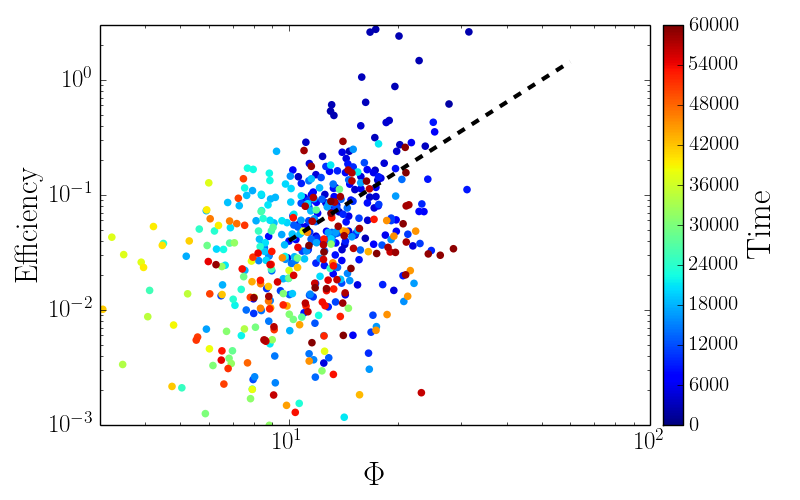}
\caption{Efficiency of jet as a function of magnetic flux
  parameter for runs (top to bottom)  \texttt{r302a9mad},
  \texttt{r302a9madhb}, and \texttt{r300a9}. Points 
correspond to snapshot data taken every $\Delta t =500$. Colors denote
the time. Jet power was calculated at $R=15$. The dashed lines show
the expected
$\varphi^2$ dependence.}
\label{f.eff}
\end{figure}

To study how the amount of energy outflowing from the system
depends on the magnetic flux threading the horizon, we
took for each simulation snapshots spaced every $\Delta t=50$, 
and extracted the accretion efficiency, defined as,
\be 
\label{eq.eta}
\eta = \frac {
\int_0^{\pi}(\rho u^r+T^r_t+R^r_t)\sqrt{-g}\, {\rm d}\theta}
{\int_0^{\pi}\rho u^r\sqrt{-g}\, {\rm d}\theta},
\ee 
where the integrals are taken at $R=15$ (to avoid the departure
from the constant energy flux close to the BH). Figure~\ref{f.eff}
shows this quantity against the magnetic flux at the horizon parameter
$\varphi$ (Eq.~\ref{eq.Phi}). Colors of points denote the time
each data point corresponds to. For the two strongly magnetized
runs (\texttt{r302a9mad} and \texttt{r302a9madhb}, first two panels) 
the points cluster along $\propto \varphi^2$ line (but for the very initial 
dark blue points), what is consistent with the prediction of the 
Blandford-Znajek power \citep[see e.g.,][]{penna+bz}. The vertical spread of points below
that line is significantly smaller for purely dipolar run
\texttt{r302a9madhb}.
The bottom panel corresponds to simulation  \texttt{r300a9} which
has never accumulated significant magnetic field. As the result 
the $\varphi$ parameter never significantly exceeds $\varphi=20$, and
the accretion efficiency remains relatively low, but still larger
than for the corresponding run \texttt{r300a0}
(Fig.~\ref{f.enflux.sane}).
Because the energy output due to the Blandford-Znajek process
is often in this case comparable with the accretion-related 
(which for a thin disk with $a_*=0.9$ equals $\eta_0=0.156$), the spread of points in
the bottom panel of Figure~\ref{f.eff} is 
significant.

\begin{figure}
\includegraphics[width=1.0\columnwidth]{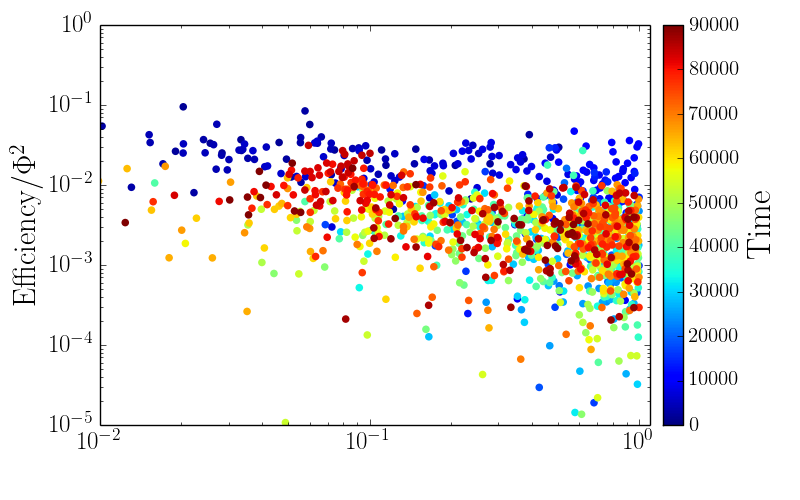}\vspace{-.45cm}
\includegraphics[width=1.0\columnwidth]{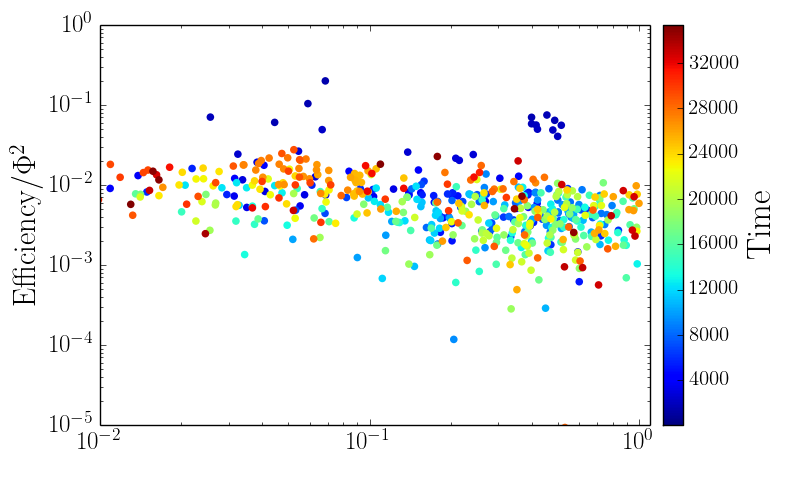}\vspace{-.45cm}
\includegraphics[width=1.0\columnwidth]{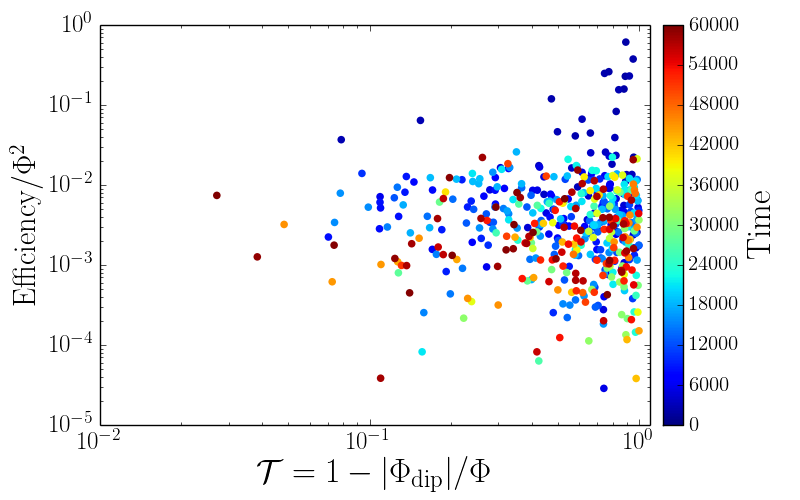}
\caption{Jet power reduced by square of the magnetic flux as a
  function of the dipolar parameter ${\cal T}=1-\varphi_{\rm dip}/\varphi$  for runs
  (top to bottom)  \texttt{r302a9mad}, \texttt{r302a9madhb}, and \texttt{r300a9}.
}
\label{f.effquad}
\end{figure}

To assess if the energy outflow depends on the topology
of the magnetic field crossing the horizon, we plot 
in Figure~\ref{f.effquad} the
energy output normalized by $\varphi^2$ (which quantity
has a flat distribution) as a function of quantity
parametrizing the field topology, which we define as,
\be
\label{eq.topopar}
{\cal T}=1-|\varphi_{\rm dip}|/\varphi.
\ee
Purely dipolar and quadrupolar fields give ${\cal T}=0$ and ${\cal
  T}=1$, respectively. Figure~\ref{f.effquad} suggests that
there is a weak dependence of the energy outflow
on the topology of the magnetic field. Dipolar magnetic field
on average produces stronger energy output, but purely 
quadrupolar field results in significant energy outflow as well.

\subsubsection{Luminosity}
\label{s.luminosities}

In the standard model of a thin accretion disk,
the whole gravitational energy liberated by viscosity
is immediately transported by diffusion to the disk surface
and emitted as radiation. As a result, the total (integrated
over radius and frequency) radiative flux corresponds
to the thin disk efficiency which depends only on the 
binding energy of gas at the disk inner edge (Eq.~\ref{e.eta}).
The luminosity of a thin disk is therefore given by,
\be
\label{e.lum1}
L_{\rm thin}=\eta_0\dot M c^2.
\ee
Introducing the Eddington accretion rate (Eq.~\ref{e.medd}),
we also have,
\be
\label{e.lum2}
\frac{L_{\rm thin}}{L_{\rm Edd}}=\frac{\dot M}{\Medd}.
\ee

Thin disks are radiatively efficient. According to the 
standard analytical modeling, it is no longer true
when accretion rate is large, and the advective contribution
to cooling is important. Such disks are called ``slim'' \citep{abra88},
and their efficiency significantly drops down because 
photons can effectively be advected on the BH, instead of
being radiated away.

Simulations of radiative, turbulent accretion disks show that the
above picture is over-simplified. Because of the turbulent structure
of the disk, other means of energy transport, e.g., magnetic buoyancy,
convection, may be important.  Energy is also carried out by jets or
winds, which is not taken into consideration in analytical disk
models.  Finally, analytical models rely on separating the radial and
vertical structure equations, but this separation becomes questionable
when the disk is geometrically thick.  Detailed study of the energy
transport inside global accreting disks is extremely interesting, but
is not within the scope of this paper. In this section we limit
ourselves to discuss various measures of the disk luminosities.

The most fundamental measure of the amount of energy
extracted by accretion is the flux of ``energy at infinity'',
i.e., the flux of kinetic, internal, and radiative 
energies that can be deposited into the interstellar
or intergalactic medium. The total flux of ``energy at infinity'' is,
\be
\label{e.lum3}
L_{\rm tot}=2\pi\int_0^\pi \left(\rho u^r + T^r_t+R^r_t\right)\sqrt{-g}\,{\rm d}\theta,
\ee
and the corresponding accretion efficiency,
\be
\label{e.lum4}
\eta_{\rm tot}=\eta_0\frac{L_{\rm tot}}{L_{\rm Edd}}\frac{\Medd}{\dot
  M}.  \ee Generally speaking, the total energy flux equals to the
average specific energy (Bernoulli number) of the gas falling through
the horizon. This is the quantity we considered in Section~\ref{s.accreff} (we
evaluated it at a specific radius: $R=15$).

Another measure of luminosity that is important
from the observational point of view is the radiative
luminosity, i.e., the flux of energy carried by photons
to an observer. However, extracting this quantity from 
numerical simulations like ours is not straightforward.
Ideally, one would like to know the disk equilibrium
solution in the whole region extending from the BH horizon
to the photosphere surface. As Figure~\ref{f.photospheres} shows,
the photosphere in the super-critical disks forms optically thin 
funnel near the axis, but extends far near the equatorial plane.
Infact, the location of the photosphere near the equatorial 
plane is determined by the extent of the initial torus (Fig.~\ref{f.initial}),
and the dynamics of gas ejected from its surface.
Because of the limited duration of simulations, the inflow/outflow
equilibrium is reached only within $R\lesssim 60$ near the equatorial
plane --- anything outside this radius is not consistent with
the accretion solution, and rather reflects the initial state of
each simulation. Therefore, calculating the radiative luminosity
by integrating the radiative flux over the solid angle near the
outer boundary would be meaningless. What is more, the outer radius
adopted in this work, $R_{\rm out}=5000$, is still not enough to encompass 
the whole photosphere for the highest accretion rates.

For these reasons, when calculating the radiative luminosity, we limit
the range of integration over $\theta$ to the optically thin funnel
region. Photons that entered this region are expected to reach the
observer because of low optical depth along the line of sight. In this
way, this measure of luminosity, calculated as, \be
\label{e.lum5}
L_{\rm f,rad}=2\pi\int_{\rm funnel} R^r_t\sqrt{-g}\,{\rm d}\theta,
\ee
may be interpreted as the lower limit of radiative luminosity
(photons may diffuse into the optically thin funnel from 
the optically thick wind beyond the boundaries of the domain).
For completeness, we calculate also the total energy 
flux in the funnel region,
\be
\label{e.lum6}
L_{\rm f,tot}=2\pi\int_{\rm funnel} \left(\rho u^r +
T^r_t+R^r_t\right)\sqrt{-g}\,{\rm d}\theta.  \ee We evaluate both of
these quantities at the limiting radius of steady state $R=R_f$ (see
Table~\ref{t.luminosities} for values).  The corresponding efficiencies are calculated
in a similar way to Eq.~\ref{e.lum4}. Knowing the luminosity inside
the optically thin region we may also calculate the average radiative
flux there in units of the isotropic Eddington flux, $F_{\rm
  Edd}=L_{\rm Edd}/4\pi R^2$, \be
\label{e.lum65}
\frac {F_{\rm f,rad}}{F_{\rm Edd}}=\frac{1}{1-\cos \theta_{\rm f}}\frac {L_{\rm f,rad}}{L_{\rm Edd}},
\ee
where $\theta_{\rm f}$ is the opening angle of the funnel.

These three measures of luminosity, together with the measure of the
radiative flux in the funnel for the same BH spin and accretion rate
are given in Table~\ref{t.luminosities}. The numbers in brackets give
the corresponding efficiencies. For comparison, the radiative
luminosities and efficiencies of the analytical thin and slim disk
models are also listed.

\begin{table*}
\centering
\caption{Accretion disk luminosities}
\label{t.luminosities}
\begin{tabular}{lcccccccccc}
\hline
\hline
Model & $\dot M$ & $L_{\rm tot}$ ($\eta_{\rm tot}$) & $L_{\rm f,tot}$ ($\eta_{\rm f,tot}$) & $L_{\rm  f,rad}$ ($\eta_{\rm  f,rad}$) & $F_{\rm f,rad}$ &
 \quad\quad &$L_{\rm thin,rad}$ ($\eta_{\rm thin,rad}$) & $L_{\rm slim,rad}$ ($\eta_{\rm slim,rad}$) &\quad & $R_{\rm f}$ \\
\hline
\texttt{r302a0} & 559 & 456 (0.046)    & - & - & -  && 559 (0.057)  & 12.1 (0.001)&& - \\
\texttt{r3015a0} & 73.1 & 55.4 (0.043) & - & - & - && 73.1 (0.057) & 9.0 (0.007)&& -\\
\texttt{r301a0} & 24.3 & 19.0 (0.045)  & 5.39 (0.013)   & 1.50 (0.004) & 39.1 && 24.3 (0.057) & 6.7 (0.016)&& 4800 \\
\texttt{r300a0} & 9.6 & 7.31 (0.043)   & 3.45 (0.020)   & 2.11 (0.013) & 7.19 && 9.6 (0.057)  & 4.6 (0.027)&& 2000 \\
\texttt{r299a0} & 2.1 & 1.81 (0.049)   & 1.80 (0.049)   & 1.65 (0.045) & 6.53 && 2.1 (0.057)  & 1.8 (0.049)&& 1200 \\
\texttt{r302a9madhb} & 3050 & 5080 (0.260)    & 527 (0.027)   & 240 (0.012) & 5100 && 3050 (0.156)  & 12.9 (0.0007)&& 4700 \\
\texttt{r302a9mad} & 2060 & 3290 (0.249)      & 176 (0.013)   & 110 (0.008) & 19000 && 2060 (0.156)  & 12.7 (0.001)&& 4700 \\ 
\texttt{r300a9} & 11.9 & 7.93 (0.104)        & 4.33 (0.057)  & 2.60 (0.034) & 37.6  && 11.9 (0.156)  & 5.2 (0.068)&& 3200 \\
\hline
\hline
\multicolumn{11}{l}{Accretion rates, luminosities, and fluxes are given in the Eddington units; numbers in brackets give the accretion 
efficiency, $\eta=L/\dot Mc^2$,}\\
\multicolumn{11}{l}{$L_{\rm tot}$ - total energy flux measured at
  $R=15$,}\\
\multicolumn{11}{l}{$L_{\rm f,tot}$, $L_{\rm f,rad}$ - respectively total and
  radiative energy flux measured at
  $R_{\rm f}$ within the optically thin funnel,}\\
\multicolumn{11}{l}{$F_{\rm f,rad}$ - average radiative flux inside the optically thin region measure at $R_{\rm f}$,}\\
\multicolumn{11}{l}{$L_{\rm thin,rad}$, $L_{\rm slim,rad}$ - radiative luminosities of a thin and slim disk \citep{sadowski-slim}, respectively, with
  given BH spin and accretion rate.}
\end{tabular}
\end{table*}

The third column in Table~\ref{t.luminosities} gives the total luminosity, $L_{\rm tot}$. 
All the simulations with a non-rotating BH show
consistent accretion efficiency, $\eta_{\rm tot}=0.043\div 0.049$
(that fraction of the accreted rest mass energy is liberated
in all other forms of energy). This number is slightly lower than the
thin disk efficiency ($\eta_0=0.057$) and indicates that
gas falls onto the BH with slightly higher specific energy than the one 
corresponding to test particles at ISCO. The efficiency
of simulated disks is significantly higher than the efficiency of
slim disks which predict significant drop in efficiency
with increasing accretion rate (down to $\eta_{\rm slim}=0.001$ for
$\sim 500\Medd$) due to the photon trapping and advection.
This fact results from the presence of outflows in the 
numerical simulations, which are ignored in the slim disk
approach.

The total accretion efficiencies for the runs
with spinning BHs are significantly higher because
of the extra source of energy --- the Blandford-Znajek
mechanism extracting the rotational energy of the BH.
The two runs initiated with single loop of magnetic field
(\texttt{r302a9madhb} and \texttt{r302a9mad})  have
higher efficiencies than the one initiated with multiple quadrupolar
 loops (\texttt{r300a9}) because they managed to 
accumulate more magnetic flux at the horizon (Fig.~\ref{f.phivst}).
The exact values of these efficiencies depend on the average
magnetic flux parameter in the period of time used for 
averaging ($\varphi\approx 25$ for \texttt{r302a9madhb} and \texttt{r302a9mad}).

The fourth and fifth columns of Table~\ref{t.luminosities} show
the total and radiative efficiencies measured by integrating
the corresponding energy fluxes inside the optically thin
funnel at some large radius $R_{\rm f}$ (given in the last column). 
The two simulations with the highest accretion rates
were so optically thick that there was virtually no photosphere
inside the simulation domain. The other three runs
with a non-rotating BH show that the larger the accretion rate
(and narrower the funnel), the smaller fraction of the
total and radiative energy fluxes escapes through the funnel. 
The remaining part flows out in the optically thick region of
outflow, and may ultimately cross the photosphere (outside
of the simulation domain), or got absorbed by the gas increasing 
its thermal or kinetic energies.
In the case of the run \texttt{r299a0}, accreting at the lowest rate ($2.1\Medd$),
nearly all the energy escapes through the funnel, mostly in
form of the radiative flux. The simulations with spinning BHs show
that the efficiencies of the energy fluxes in the funnel are significantly
lower than the net efficiencies ($\eta_{\rm tot}$), reflecting
the fact that most of the energy is bound in an outflowing optically thick wind.

Although the radiative luminosites in the funnel are moderate for the
non-rotating BH disks, the effective flux of radiation in the funnel
(sixth column in Table~\ref{t.luminosities}) significantly exceeds the
Eddington value because of the small solid angle covered by the
radiative flux there.  Already for simulation \texttt{r299a0} ($\dot M=2.1\Medd$)
the
effective flux exceeds the isotropic Eddington flux more than six
times.  The fluxes are even higher for the simulations with a spinning
BH, where the radiative flux in a narrow funnel is amplified by
spin-related extra energy outflow.

\begin{figure}
\includegraphics[width=1.0\columnwidth]{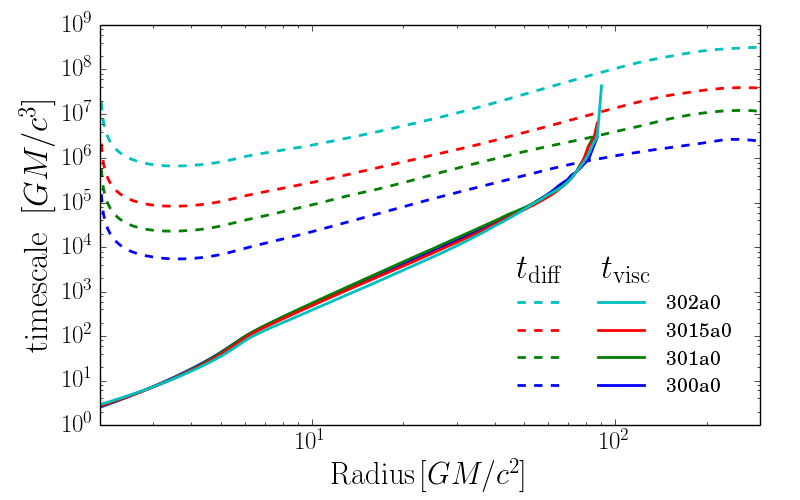}
\caption{Viscous and diffusive timescales for the four super-critical disks.}
\label{f.timescales.sane}
\end{figure}

Very large optical depths would
imply effective photon trapping. In Figure~\ref{f.timescales.sane}
we compare the viscous timescale,
\be
\label{e.lum7}
t_{\rm vis}=-\frac{R}{u^r},
\ee
where $u^r$ is the average inflow velocity,
with the photon (vertical) diffusion timescale,
\be
\label{e.lum8}
t_{\rm diff}\approx 3\tau_{\rm tot}H,
\ee
where $\tau_{\rm tot}$ and $H$ are the total vertical optical depth, and disk 
density scaleheight, respectively. 
The fact that the viscous timescale is significantly shorter
 suggests that
most of the photons inside the disk are trapped and
effectively advected inward (the solutions are advection-dominated).
 Indeed, Figure~\ref{f.enflux.sane} shows that
the net trapping radius for photons is located at $R\approx 20$.
Despite effective photon trapping, the disk is able
to liberate significant fraction of the accreted rest mass energy.
Therefore, the energy dissipated by viscosity must be efficiently
trasported outward by other processes. 
Outflows ejected by magnetic fields most likely play a crucial role. Detailed investigation
of the means of energy transport will be done in one of the following works.

\subsection{Unstable thin disks}
\label{s.unstable}

In previous sections we discussed simulations
of disks around both rotating and non-rotating BHs,
accreting at rates exceeding the Eddington limit. The ultimate accretion rate of a given 
disk was determined by the density of the initial
equilibrium torus. It is therefore straightforward
to attempt to simulate disks accreting at lower
 rates, comparable and lower than the
Eddington limit. One would expect, based
on the standard thin disk theory, that such disks 
would be thinner, colder, and accreting with
lower velocities. Simulating them would be also
more challenging --- thinner disks require smaller
grid cells near the equatorial plane what suppresses
the time step, and lower radial velocities imply
longer viscous timescales, requiring
longer, more expensive simulations. What is more,
once the accretion rate drops below $\sim 5\Medd$ 
\citep{sadowski.phd}, disk solutions are expected
to leave the slim disk branch, and enter the
thermally unstable regime of radiation pressure
dominated, radiatively efficient disks.

To our knowledge, so far only \cite{ohsuga11}
attempted to simulate such thin accretion disks
with an MHD code. However, the thin disk model
described there was not supported by 
dissipative heating --- the lack of resolution 
suppressed the MRI, and the disk collapsed because
of insufficient heating rate. 

Simulating thin accretion disks
is for these reasons very challenging and 
expensive. It is clear that in the near future
no 3D MHD simulation could even approach simulating 
disks with accretion rates $\dot M\sim 0.1 \Medd$.
This may be feasible, however, in axisymmetry
with sub-grid dynamo, which allows for long
lasting simulations. In this section we describe
our first attempt to resolve sub-critical, geometrically
thin disks.

\begin{figure}
\includegraphics[width=1.0\columnwidth]{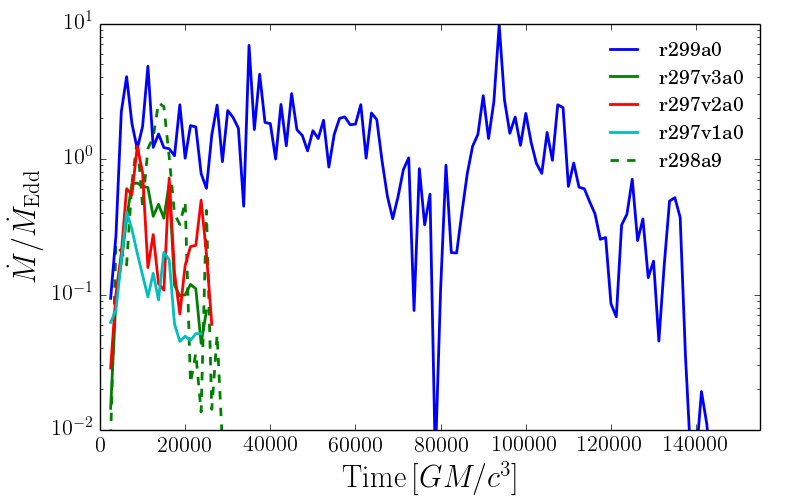}
\caption{Accretion rate history for the unstable thin disks.}
\label{f.mdotvstcollapse}
\end{figure}

We performed five simulations initiated with 
torus densities lower than for the simulations described
so far (Table~\ref{t.models}). Four of them 
were simulated with a non-rotating BH, and one 
with BH spin $a_*=0.9$. Figure~\ref{f.mdotvstcollapse}
shows the accretion rate history for these 
simulations. All runs but for \texttt{r299a0}
quickly ceased accreting gas after a short
(lasting $\sim10,000\,GM/c^3$) episode of efficient
accretion with $\dot M\sim 0.1\div 1.0 \Medd$.
Only run \texttt{r299a0} managed to accrete 
gas consistently for a longer period of time.
Until $t=60,000$, the accretion rate oscillated
around $2.1 \Medd$. Afterwards, it went down,
only to come back to the original value around
$t=90,000$, and finally showed rapid decrease and
ceased at $t=140,000$. 

\begin{figure}
\includegraphics[width=1.0\columnwidth]{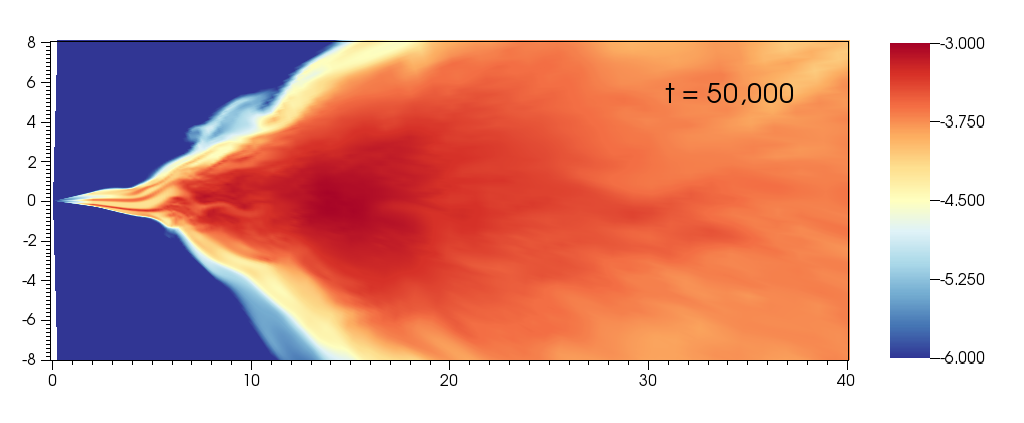}\vspace{-.45cm}
\includegraphics[width=1.0\columnwidth]{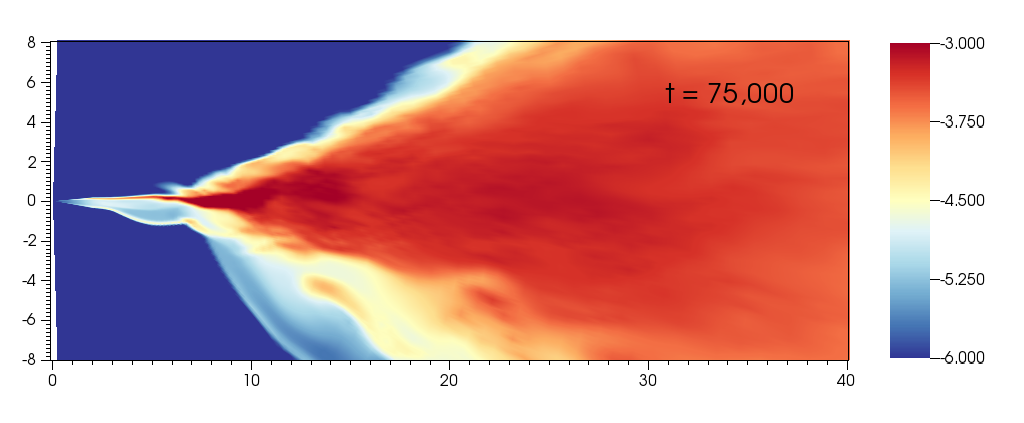}\vspace{-.45cm}
\includegraphics[width=1.0\columnwidth]{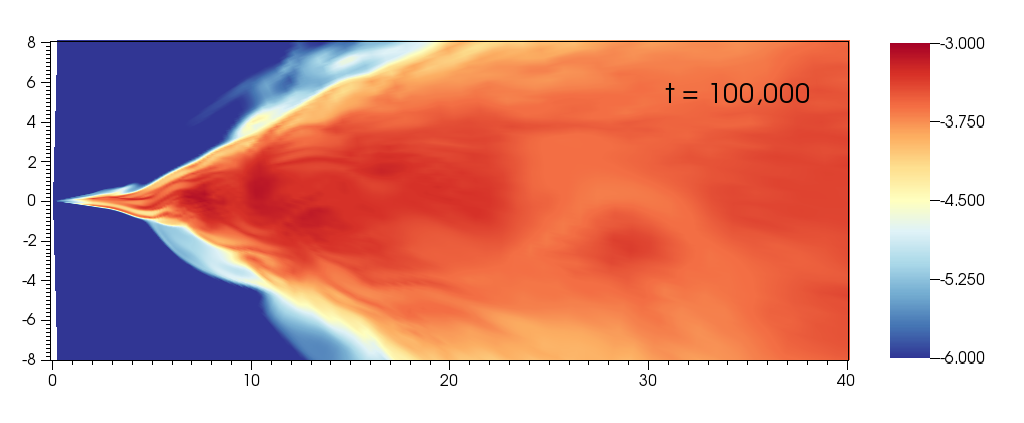}\vspace{-.45cm}
\includegraphics[width=1.0\columnwidth]{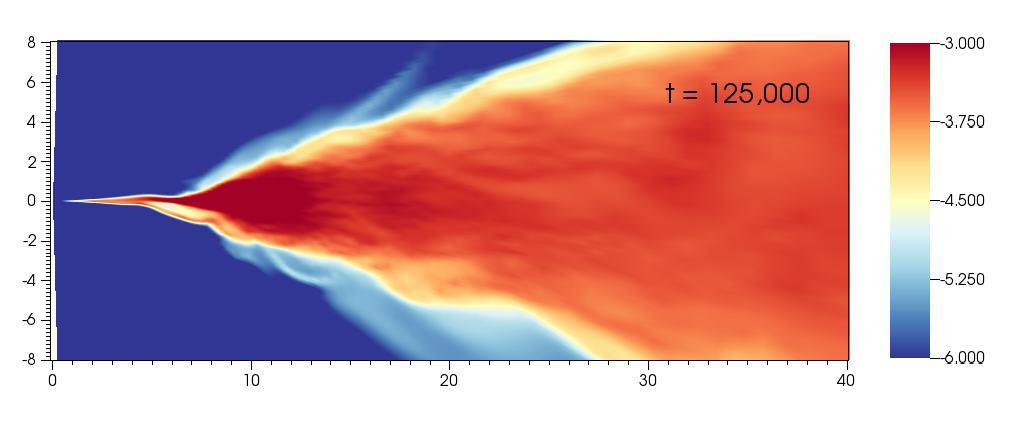}\vspace{-.45cm}
\includegraphics[width=1.0\columnwidth]{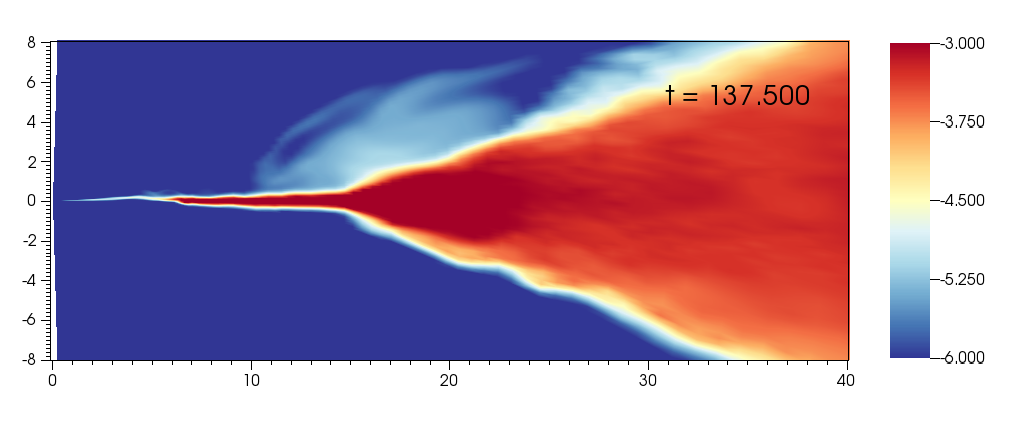}\vspace{-.45cm}
\includegraphics[width=1.0\columnwidth]{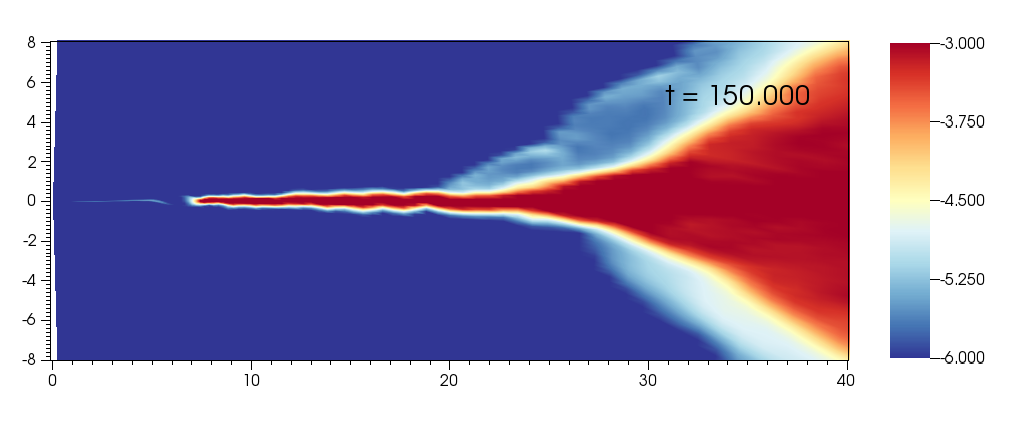}
\caption{Sequence of snapshots of logarithms of density for
run \texttt{r299a0} showing the collapsing disk. The snapshots
correspond to (top to bottom): $t=50,000$, $75,000$, $100,000$, $125,000$,
$137,500$, and $150,000$.}
\label{f.299collapse}
\end{figure}

Figure~\ref{f.299collapse} shows snapshots of
density in run \texttt{r299a0} at various moments
of time. The topmost panel corresponds to the episode
of continuous accretion at $t<60,000$, and shows
a relatively thin (density scaleheight at $H/R\approx0.2$), 
turbulent accretion disk. At $t=75,000$ (second panel),
corresponding to the first dip in the accretion rate,
the innermost part of the disk was significantly thinner
and denser. At $t=100,000$, when the accretion rate was 
again consistent and around $2\Medd$, the disk looked exactly
as in the topmost panel. This was followed by another decrease
in accretion rate, and corresponding compression of the innermost region
($t=125,000$). This time, however, the disk did not manage
to recover its original state, and, as the two bottom-most panels of 
Figure~\ref{f.299collapse} show, collapsed to a very thin and high
density state. At that point, the vertical resolution, although
already extremely high, was not enough to sustain
turbulence, MRI ceased, and accretion stopped.

\begin{figure}
\includegraphics[width=1.0\columnwidth]{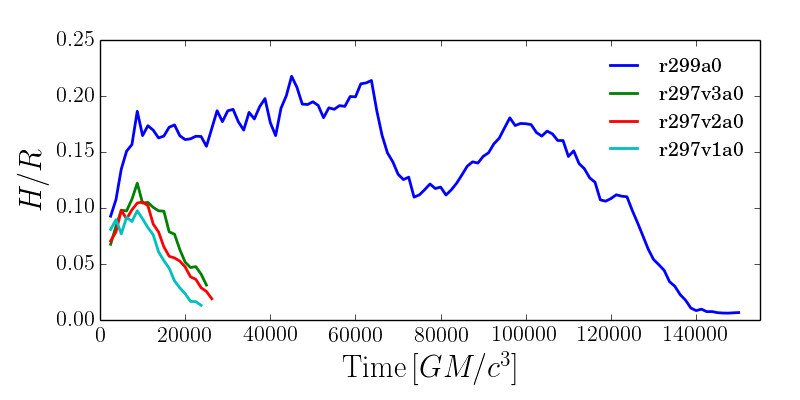}\vspace{-.53cm}
\includegraphics[width=1.0\columnwidth]{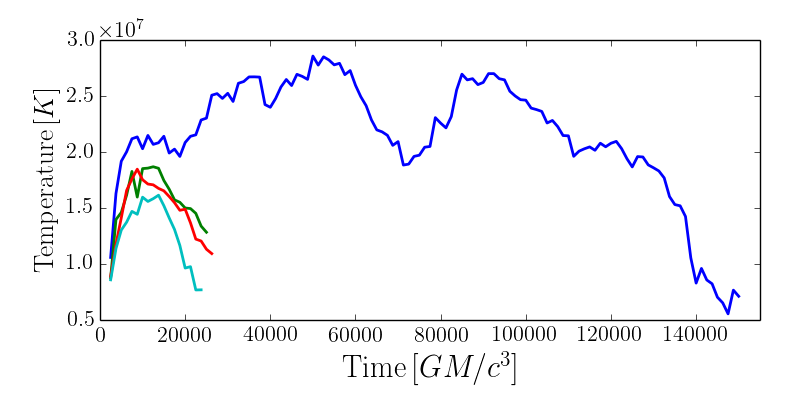}
\caption{Changes with time of disk $H/R$ ratio (top panel) and mean temperature (bottom)
measured at radius $R=15$ for the collapsing thin disk models with $a_*=0.0$.}
\label{f.doublecollapse}
\end{figure}

Figure~\ref{f.doublecollapse} shows the evolution of 
disk thickness and mean temperature at fixed radius 
$R=15$ for the four collapsing runs around non-rotating BHs.
All the runs show initial increase of thickness and temperature
before $t=10,000$ from values intrinisic to the initial torus,
towards the values corresponding to a turbulent, accreting flow.
Only run \texttt{r299a0} manages to sustain that state for
significant time. The temperature and disk thickness 
are correlated with each other for all the runs. The evolution of 
run \texttt{r299a0} suggests that drop in temperature preceeds
decrease of disk thickness. This fact suggests, that disk thickness
adjusts to the pressure in the disk to maintain vertical 
hydrostatic equilibrium. Ultimately, run \texttt{r299a0} shows
runaway cooling and collapses, as the other runs did much earlier.

\begin{figure}
\includegraphics[width=1.0\columnwidth]{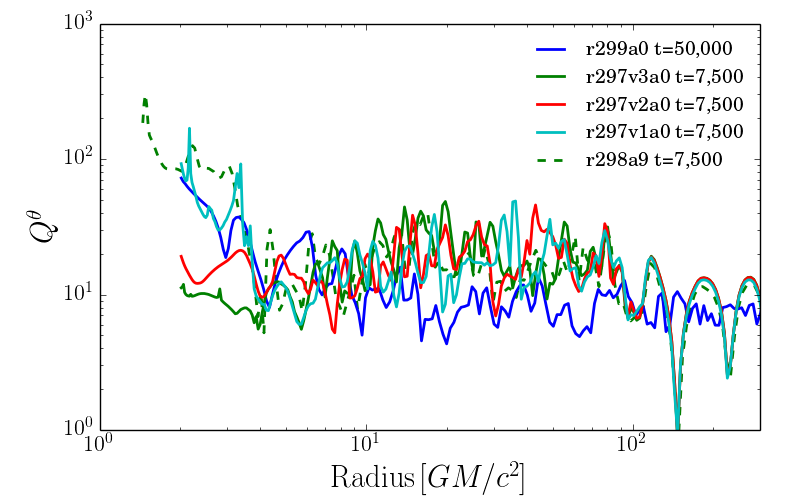}
\caption{Vertically averaged MRI resolution paratemeter $Q^\theta$ for
the unstable disks at $t=50,000$ (for \texttt{r299a0}) or $t=7,500$ for the other runs.}
\label{f.qtheta.collapse}
\end{figure}

Such runaway cooling could take place because of
insufficient resolution leading to not resolving
MRI, and effectively suppresing the heating rate.
To reject this possibility we show the MRI
resolution parameter $Q^\theta$ (Eq.~\ref{eq.qtheta}) 
in Figure~\ref{f.qtheta.collapse} at times directly 
preceding the collapse for each of the five runs.
Values of $Q^\theta$ in the innermost ($R<10$) region,
where the collapse begins, fall in range $7\lesssim Q^\theta \lesssim 30$,
and suggest that MRI is properly resolved there. To
further verify that our choice of resolution is enough
we performed two additional simulations similar to 
run \texttt{r298a0} but with nearly doubled the resolution 
in the polar angle for one run, and in radius for the other.
They both showed exactly the same behavior.

The accretion rates of these five runs (Fig.~\ref{f.mdotvstcollapse})
suggest that they belong to the thermally unstable branch of
solutions, outside the region where advective cooling can act to stabilize against thermal collapse.  
For a radiation pressure dominated radiatively efficient disk with $\alpha$-viscosity, the heating rate
($Q^+\propto \alpha p_0 H\propto p_0^2$)
has a steeper dependence on the midplane total 
pressure $p_0$ (at fixed surface density and $\alpha$ viscosity parameter)
than the radiative cooling rate ($Q^-\propto T^4\propto p_0$). This
fact inevitably leads either to a local runaway collapse or expansion
of the disk, and may trigger a large scale evolution of 
the disc structure (limit cycle) similar to the one seen in 
dwarf nova outbursts \citep[see][for a review]{L01}. The fact that run \texttt{r299a0}
shows significantly less rapid collapse may be explained by the
fact that its accretion rate is very close to the
turnover point at $\sim 5\Medd$. We believe that this mechanism is behind
the collapse observed in our simulations. More detailed study
of the energy balance is beyond the scope of this paper.

One of the major puzzles in our understanding of BH accretion disks is
the fact that the thermal instability suggested by the $\alpha$-viscosity
model (and seen in our simulations) does not take place for most, if not
all of the BH binaries, which show steady lightcurves even for luminosities
corresponding to the unstable range of accretion rates. Most tempting
explanation would be to say that the assumption that $\alpha$-viscosity
properly described the viscous stress-energy tensor arising from MRI-driven
turbulence is wrong. This could be verified by performing local simulations
of vertically stratified, radiation-pressure dominated gas. Initial 
work by \cite{hirose09b}, performed with the flux-limited diffusion radiative
transfer scheme, suggested that turbulent flows in this
regime are indeed thermally stable. However, \cite{jiang+13} repeated their 
study using improved radiative closure, and showed that the stability
is affected not only by the way the radiation is treated, but also by
the size of the box considered. For their most reliable setup, the 
disk annuli was shown to be thermally unstable. Therefore, 
the puzzle remains open.

One has to keep in mind, that although the dynamo-sustained turbulence that
characterizes our axisymmetric accretion disks has mean properties
similar to the 3D turbulence, some aspects of it, e.g., response to
compression, may be different from properties of 3D turbulence. Therefore,
to study thermal stability self-consistently, one has to 
ultimately perform 3D simulations. Our model may provide
only limited insights into this issue.

\section{Summary and discussion}
\label{s.summary}

In this paper we have introduced a new sub-grid prescription 
for the magnetic dynamo that allows axisymmetric simulations 
of accretion disks to be run for arbitrarily long times. 
The dynamo model generates a weak correction to the global poloidal field, and drives it
towards the properties of a 3D saturated turbulent state, namely,
the proper value of the magnetic field angle (Eq.~\ref{eq.bangle})
and magnetic to total pressure ratio (Eq.~\ref{e.betaprime}).
We have implemented the dynamo model into a radiative, general
relativistic, ideal MHD code \texttt{KORAL}. We also introduced 
a viscous correction to the M1 radiative closure scheme which
prevents artificial shocks of the radiation field near the rotation
axis.

We performed a
set of twelve axisymmetric, global simulations of optically thick
accretion disks. From among them, seven evolved into quasi-stationary,
super-Eddington accretion disks, and the remaining five, corresponding
to disks with the lowest, mostly sub-Eddington, accretion rates, collapsed
because of runaway cooling.

Our study shows that:

(i) The proposed sub-grid dynamo prescription effectively prevents the
poloidal magnetic field from decaying, and successfully drives the
properties of the magnetic field towards the prescribed
characteristics corresponding to the saturated state as inferred from
3D shearing sheet simulations. However, the agreement is not perfect inside the
innermost stable orbit, where the radial velocities and magnetization
are the largest.

(ii) The viscous correction to the M1 closure described in Appendix~\ref{ap.viscosity} successfully smooths
the radiative field near the polar axis. Nevertheless, M1 closure 
still has its own limitations, and a more detailed study is required
to assess how important the choice of closure scheme is for the dynamics of 
an accretion flow.

(iii) Optically thick accretion can take place in a quasi-stationary
way for accretion rates $\dot M\gtrsim 2\Medd$. Simulations
initiated with torii designed to provide lower accretion rates 
exhibited a run-away cooling leading to a disk collapse.
The fact that the MRI seemed to be resolved before the collapse, and
the critical accretion rate below which the collapse occurs ($\dot M_{\rm crit}\approx 2\Medd$ which is where the
transition between the middle, radiation pressure-dominated, and the top, advection-dominated, branches is expected)
 suggest that this is the result of a thermal 
instability. However, more detailed analysis is required to
confirm this hypothesis. One should also keep in mind that although 
our dynamo-sustained magnetic field has similar properties to saturated 
fields obtained in 3D studies, our disks are turbulent only in
the poloidal plane. Ultimately 3D simulations are needed for a 
self-consistent study of the onset of thermal instability.

(iv) The simulations of super-Eddington accretion flows we performed suggest that the radius outside
which outflow in a wind is present ($R\approx 20$ and $R\approx 10$ for
$a_*=0$ and $0.9$, respectively) is smaller than for corresponding
simulations of radiatively inefficient, optically thin disks \citep{narayan+12,sadowski+outflows}.
This fact suggest that winds are more effectively driven out
in radiation-pressure supported optically thick disks.
Within the range of accretion rates explored in our simulations with $a_*=0$ 
($2\lesssim \dot M/\Medd \lesssim 500$) the relative power
of the wind seems to be independent of the accretion rate.
Extending this trend down would suggest that similar outflows may also occur
in thin disks accreting at low rates and affect their dynamics.

(v) The total efficiency of super-critical accretion on a non-rotating
BH is consistent among all four simulations we performed and
corresponds to extraction of $\sim 4\%$ of the accreted rest mass
energy, independent of the accretion rate.  This is slighly less than
the efficiency corresponding to a classical thin accretion disk around
a non-rotating BH ($\eta_0=0.057$), but is much higher than the
efficiencies predicted by the slim disk model
\citep[e.g.,][]{sadowski.phd}.  This extra efficiency is related to
the existence of outflows ejected by accretion.  Because of limited
duration and limited size of the simulation domain, we are not able
to give exact radiative luminosities, and instead can only provide a
lower limit (corresponding to the luminosity from inside the optically
thin region of the domain).  In most cases, this radiative luminosity
estimate comprises only a small fraction of the total flux of
outflowing energy.


(vi) Accretion disks around spinning BHs extract rotational
energy from the BH, and this energy extraction process constitutes
a large fraction of the total energy efficiency for the system.
The energy extraction rate attributed to the Blandford-Znajek 
mechanism has comparable efficiency to that measured in 3D simulations
of optically thin advection dominated accretion flows (ADAFs). In this work,
we are not able to study this in detail --- axisymmetric simulations are
limited to relatively weak magnetic fields at the BH, and thus we cannot
resolve the magnetically arrested state (MAD).

(vii) The sub-grid dynamo model generates a turbulent field
which overwrites the topology of the initial magnetic field (assuming weak starting fields). 
As a result, the magnetic field threading the horizon
is not a simple dipole. It often changes polarity, and may have 
multipolar structure owing to the tangled nature of turbulent magnetic fields. 
Despite the complex multipolar nature of the magnetic field, the end result 
still seems to be efficient energy extraction from the BH at levels comparable to
the prediction from the BZ mechanism.  On average, the rapidly varying configuration 
has only slightly lower efficiency than a structured purely dipolar field.

(viii) Super-critical accretion is optically thick, and the photosphere
may be located very close to the polar axis. For accretion rates
exceeding $\sim 20\Medd$, there is no optically thin region at all
inside $R=5000$. In all cases, gas gradually rarifies, the funnel opens
and photons diffuse into it from the optically thick region of outflowing
gas.

(ix) The radiative luminosities in the funnel for
the runs with a non-rotating BH are relatively low ($1.5\div2.11L_{\rm Edd}$)
despite significantly super-Eddington accretion rates. However, the corresponding
radiative fluxes near the polar axis are extremely high, reaching
$\sim 1000$ times the Eddington flux, $F_{\rm Edd}=L_{\rm Edd}/4\pi R^2$.
It shows, that the radiative emission from a super-critical accretion 
disk is strongly collimated along the axis. If only a source with
moderate accretion rate is observed down the funnel, the apparent
luminosity of such a source will be orders of magnitude higher.
To assess the relevance of this effect to the ultraluminous X-ray sources,
detailed radiative transfer modeling is required.

\section{Acknowledgements}

AS and RN were
supported in part by NSF grant AST1312651 and NASA grant NNX11AE16G.
AT was supported by NASA
through Einstein Postdoctoral Fellowship grant number PF3-140115
awarded by the Chandra X-ray Center, which is operated by the
Smithsonian Astrophysical Observatory for NASA under contract
NAS8-03060.
We also acknowledge computational support from NSF via XSEDE resources
(grant TG-AST080026N to RN and AS, and grant TG-AST100040 to AT), and
from NASA (to RN and AS) via the High-End Computing (HEC) Program
through the NASA Advanced Supercomputing (NAS) Division at Ames
Research Center.
 
\bibliographystyle{mn2e}

{\small
\begin{thebibliography}{}



\bibitem[Abramowicz et al.(1988)]{abra88} Abramowicz, M.~A., Czerny,
  B., Lasota, J.~P., \& Szuszkiewicz, E.\ 1988, \apj, 332, 646

\bibitem[Anninos et al.(2005)]{anninosetal05} Anninos, P., Fragile, 
P.~C., \& Salmonson, J.~D.\ 2005, Cosmos++: Relativistic
Magnetohydrodynamics on Unstructured Grids with Local Adaptive
Refinement, \apj, 635, 723 


\bibitem[Blackman, Penna, 
\& Varni{\`e}re(2008)]{blackman+08} Blackman, E.~G., Penna, R.~F., \& Varni{\`e}re, P.\ 2008, New Astronomy, 13, 244 



\bibitem[Blaes et al.(2007)]{blaes2007} Blaes, O., Hirose, S., 
\& Krolik, J.~H.\ 2007, \apj, 664, 1057 

\bibitem[Blaes et al.(2011)]{blaes2011} Blaes, O., Krolik, J.~H., 
Hirose, S., \& Shabaltas, N.\ 2011, \apj, 733, 110 

\bibitem[Blandford 
\& Znajek(1977)]{bz} Blandford, R.~D., \& Znajek, R.~L.\ 1977, \mnras, 179, 433 



\bibitem[Brandenburg et al.(1995)]{brandenburg+95} Brandenburg, A., Nordlund, A., Stein, R.~F., \& Torkelsson, U.\ 1995, \apj, 446, 741 




\bibitem[Brandenburg(2001)]{brandenburg-01} Brandenburg, A.\ 2001, 
\apj, 550, 824 

\bibitem[Bucciantini 
\& Del Zanna(2013)]{buccidelzanna+13} Bucciantini, N., \& Del Zanna, L.\ 2013, \mnras, 428, 71 





\bibitem[Chandrasekhar(1950)]{chandra-50} Chandrasekhar, S.\ 1950, 
Oxford, Clarendon Press, 1950.,  




\bibitem[Cowling(1933)]{cowling-33} Cowling, T.~G.\ 1933, \mnras, 
94, 39 




\bibitem[Del Zanna et 
al.(2007)]{delzannaetal07} Del Zanna, L., Zanotti, O., Bucciantini,
N., \& Londrillo, P.\ 2007,  \aap, 473, 11 

\bibitem[De Villiers et al.(2003)]{devilliersetal03} De Villiers, J.-P., 
Hawley, J.~F., \& Krolik, J.~H.\ 2003,  \apj, 599, 1238 


\bibitem[Dubroca \& Feugeas(1999)]{dubrocafeugeas99} Dubroca, B., \& Feugeas, J. L. 1999, CRAS, 329, 915

\bibitem[Duez et al.(2004)]{duez+04} Duez, M.~D., Liu, Y.~T., Shapiro, S.~L., \& Stephens, B.~C.\ 2004, \prd, 69, 104030 




\bibitem[Gammie et al.(2003)]{gammie03} Gammie, C.~F., McKinney, 
J.~C., \& T{\'o}th, G.\ 2003, \apj, 589, 444 

\bibitem[Guan  et al.(2009)]{guan+09} Guan, X., Gammie, C.~F., Simon, J.~B., \& Johnson, B.~M.\ 2009, \apj, 694, 1010 





\bibitem[Hawley et al.(2011)]{hawley+11} Hawley, J.~F., Guan, X., 
\& Krolik, J.~H.\ 2011,  \apj, 738, 84 

\bibitem[Hawley et al.(2013)]{hawley+13} Hawley, J.~F., Richers, S.~A., Guan, X., \& Krolik, J.~H.\ 2013, \apj, 772, 102 


\bibitem[Hirose et al.(2009a)]{hirose09b} Hirose, S., Krolik, J. H. \&
  Blaes, O.\ 2009a, \apj, 691, 16

\bibitem[Hirose et al.(2009b)]{hirose09a} Hirose, S., Blaes, O., 
\& Krolik, J.~H.\ 2009b, \apj, 704, 781 



\bibitem[Jiang et al.(2012)]{jiang+12} Jiang, Y.-F., Stone, 
J.~M., \& Davis, S.~W.\ 2012, \apjs, 199, 14 

\bibitem[Jiang, Stone, 
\& Davis(2013)]{jiang+13} Jiang, Y.-F., Stone, J.~M., \& Davis, S.~W.\ 2013, \apj, 778, 65 

\bibitem[Jiang, Stone, 
\& Davis(2014)]{jiang+14} Jiang, Y.-F., Stone, J.~M., \& Davis, S.~W.\ 2014, arXiv:1403.6126 




\bibitem[Kawashima et al.(2009)]{kawashima+09} Kawashima, T., 
Ohsuga, K., Mineshige, S., Heinzeller, D., Takabe, H., 
\& Matsumoto, R.\ 2009, \pasj, 61, 769 




\bibitem[Komissarov(1999)]{komissarov-99} Komissarov, S.~S.\ 1999, 
\mnras, 303, 343 



\bibitem[Krolik et al.(2007)]{kro07} Krolik, J.~H., Hirose, 
S., \& Blaes, O.\ 2007, \apj, 664, 1045

\bibitem[Krolik 
\& Piran(2012)]{krolikpiran-12} Krolik, J.~H., \& Piran, T.\ 2012,  \apj, 749, 92 


\bibitem[Landau 
\& Lifshitz(1959)]{landau+fluid} Landau, L.~D., \& Lifshitz, E.~M.\ 1959, Course of theoretical physics, Oxford: Pergamon Press, 1959,  

\bibitem[Lasota(2001)]{L01} Lasota, J.-P.\ 2001, New Astronomy Reviews, 45, 
449 


\bibitem[Lasota et al.(2014)]{lasota+bz} Lasota, J.-P., 
Gourgoulhon, E., Abramowicz, M., Tchekhovskoy, A., 
\& Narayan, R.\ 2014, \prd, 89, 024041 



\bibitem[Levermore(1984)]{levermore84} Levermore, C.~D.\ 1984, 
\jqsrt, 31, 149 


\bibitem[Li 
\& Narayan(2004)]{linarayan-04} Li, L.-X., \& Narayan, R.\ 2004, \apj, 601, 414 




\bibitem[McKinney(2006)]{mckinney06} McKinney, J.~C.\ 2006, 
\mnras, 368, 1561 



\bibitem[McKinney et al.(2012)]{mtb12} McKinney, J.~C., 
Tchekhovskoy, A., \& Blandford, R.~D.\ 2012, \mnras, 423, 3083 

\bibitem[McKinney et al.(2013)]{mckinney+harmrad} McKinney, J.~C., 
Narayan, R., S{\c a}dowski, A., \& Tchekhovskoy, A.\ 2013, \mnras, submitted




\bibitem[Narayan et al.(2012)]{narayan+12} Narayan, R., S{\c a}dowski, A., Penna, R.~F., \& Kulkarni, A.~K.\ 2012,  \mnras, 426, 3241 


\bibitem[Ohsuga et al.(2009)]{ohsuga09} Ohsuga, K., Mineshige, 
S., Mori, M., \& Yoshiaki, K.\ 2009, \pasj, 61, L7 


\bibitem[Ohsuga 
\& Mineshige(2011)]{ohsuga11} Ohsuga, K., \& Mineshige, S.\ 2011, \apj, 736, 2 



\bibitem[Parker(1955)]{parker-55} Parker, E.~N.\ 1955, \apj, 122, 
293 

\bibitem[Penna et al.(2013a)]{penna-limotorus} Penna, R.~F., Kulkarni, A., \& Narayan, R.\ 2013a, \aap, 559, A116 

\bibitem[Penna et al.(2013b)]{penna+alpha} Penna, R.~F., S{\c 
a}dowski, A., Kulkarni, A.~K., \& Narayan, R.\ 2013b, \mnras, 428, 2255 

\bibitem[Penna et al.(2013c)]{penna+bz} Penna, R.~F., Narayan, 
R., \& S{\c a}dowski, A.\ 2013c, \mnras, 436, 3741 






\bibitem[Pozdnyakov et al.(1983)]{pozdnyakov+83} Pozdnyakov, L.~A., Sobol, I.~M., \& Syunyaev, R.~A.\ 1983, Astrophysics and Space Physics Reviews, 2, 189 


\bibitem[Press et al.(1986)]{numericalrecipes} Press, W.~H., Flannery, 
B.~P., \& Teukolsky, S.~A.\ 1986, Cambridge: University Press, 1986,  

\bibitem[Rybicki 
\& Lightman(1979)]{rybicki-book} Rybicki, G.~B., \& Lightman, A.~P.\ 1979, New York, Wiley-Interscience, 1979.~393 p.,  



\bibitem[S{\c a}dowski(2009)]{sadowski-slim} S{\c a}dowski, A.\ 2009, 
\apjs, 183, 171 




\bibitem[S{\c a}dowski(2011)]{sadowski.phd} S{\c a}dowski, A.\ 2011, Ph.D. Thesis, Nicolaus Copernicus
Astronomical Center, Polish Academy of Sciences, arXiv:1108.0396 



\bibitem[S{\c a}dowski et al.(2013a)]{sadowski+koral} S{\c a}dowski, 
A., Narayan, R., Tchekhovskoy, A., \& Zhu, Y.\ 2013a, \mnras, 429, 3533 




\bibitem[S{\c a}dowski et al.(2013b)]{sadowski+outflows} S{\c a}dowski, 
A., Narayan, R., Penna, R., \& Zhu, Y.\ 2013b, \mnras, 436, 3856 



\bibitem[S{\c a}dowski et al.(2014)]{sadowski+koral2} S{\c a}dowski, A., Narayan, R., McKinney, J.~C., \& Tchekhovskoy, A.\ 2014, \mnras, 439, 503 


\bibitem[Shakura \& Sunyaev(1973)]{ss73}Shakura, N. I., \& Sunyaev, R. A. 1973, A\&A, 24, 337



\bibitem[Skinner 
\& Ostriker(2013)]{skinnerostriker-13} Skinner, M.~A., \& Ostriker, E.~C.\ 2013, \apjs, 206, 21 




\bibitem[Sorathia et al.(2012)]{sorathia+12} Sorathia, K.~A., Reynolds, C.~S., Stone, J.~M., \& Beckwith, K.\ 2012, \apj, 749, 189 




\bibitem[Tajima 
\& Fukue(1998)]{tajimafukue-98} Tajima, Y., \& Fukue, J.\ 1998, \pasj, 50, 483 





\bibitem[{Tchekhovskoy} et al.(2010){Tchekhovskoy}, {Narayan}, and
  {McKinney}]{tchekh10a} {Tchekhovskoy}, A., {Narayan}, R., and
  {McKinney}, J.~C.\ 2010, \newblock {
    New Astron.}, { 15}, 749--754 

\bibitem[Tchekhovskoy 
\& McKinney(2012)]{tchekh+12} Tchekhovskoy, A., \& McKinney, J.~C.\ 2012, \mnras, 423, L55  



\bibitem[T{\'o}th(2000)]{toth-00} T{\'o}th, G.\ 2000, Journal 
of Computational Physics, 161, 605 


\bibitem[Turner et al.(2003)]{turner2003} Turner, N.~J., Stone, 
J.~M., Krolik, J.~H., \& Sano, T.\ 2003, \apj, 593, 992





\bibitem[Yuan 
\& Narayan(2014)]{yuannarayan-14} Yuan, F. \&  Narayan, R.\ 2014, Annual Reviews, submitted




\end{thebibliography}
}

\appendix

\section{Subgrid dynamo}
\label{ap.dynamo}

We describe here the model we use to introduce an effective magnetic 
dynamo mechanism
in an axisymmetric  ideal MHD simulation.
We start with the mean magnetic field evolution equation 
\citep{brandenburg-01},
\be
\label{eq.meandynamo2}
\frac{\partial\vec B}{\partial t}=\alpha \nabla \times \vec B + \eta \nabla^2 \vec B,
\ee
where $\alpha$ and $\eta$, are the dynamo and magnetic diffusivity 
coefficients, respectivelly. The first term on the left hand
side describes the dynamo effect which generates poloidal magnetic field, and
the second corresponds to the dissipation of the magnetic field. We treat the two components independently.

\subsection{$\alpha$-effect}

The
magnetic field generation term may be written in the form,
\be
\label{e.dyn1}
\pder{\vec B_{\hat p}}{t}=\alpha (\nabla \times \vec B)_{\hat p},
\ee
where we restrict ourselves to the generation
of orthonormal poloidal component $\vec B_{\hat p}$. We expect field generation 
to happen on a dynamical timescale $\Omega_{\rm K}^{-1}$, therefore
we introduce normalization factors,
\be
\label{e.dyn2}
\frac{1}{|\vec B_{\hat p}|}\pder{\vec B_{\hat p}}{t}=\alpha \frac{\Omega_{\rm K}}{\vec |J_{\hat p}|}(\nabla \times \vec B)_{\hat p},
\ee
where $J_{\hat p}=(\nabla \times \vec B)_{\hat p}$ is the poloidal
current, the magnitude of which we approximate
as,
\be
\label{e.dyn3}
|\vec J_{\hat p}|=|(\nabla \times \vec B)_{\hat p}|\approx \frac{|B_{\hat \varphi}|}{H},
\ee
with $H$ being the disk scale height at the given radius.

The poloidal magnetic field may be expressed as,
\be
\label{e.dyn4}
\vec B_{\hat p}=(\nabla \times \vec A)_{\hat p},
\ee
where $\vec A$ is the vector potential.
Putting Eqs.~\ref{e.dyn2}, \ref{e.dyn3} and \ref{e.dyn4} together,
we get,
\be
\label{e.dyn5}
\pder{}{t}(\nabla \times \vec A)_{\hat p}=
\Omega_{\rm K}\alpha (\nabla \times \vec B)_{\hat p}\frac{|\vec B_{\hat p}|H}{|B_{\hat \varphi}|}.
\ee
Now we assume that all the terms outside the brackets
are invariant with respect to curl, and get,
\be
\label{e.dyn6}
\pder{A_{\hat \varphi}}{t}=
\Omega_{\rm K}\alpha B_{\hat \varphi}\frac{|\vec B_{\hat p}|H}{|B_{\hat \varphi}|},
\ee
where we retain only the azimuthal components of the vector potential
and magnetic field inside the curl since they are the source of the poloidal 
field.

Local 3D simulations of MHD turbulence, either in the shearing
sheet approximation \citep[e.g,][]{guan+09, jiang+13}, or global \citep{sorathia+12},
have shown that the turbulent magnetic field saturates
at a magnetic field angle,
\be
\label{e.dyn7}
\xi=\frac{b^r b^\phi}{b^2}\approx\frac{|\vec B_{\hat p}|}{|B_{\hat \varphi}|}=\xi_{\rm dyn}\approx \frac{1}{4}.
\ee
We include this number into Eq.~\ref{e.dyn6} and get,
\be
\label{e.dyn8}
\pder{A_{\hat \varphi}}{t}=\alpha_{\rm dyn}
\Omega_{\rm K} H B_{\hat \varphi},
\ee
where $\alpha_{\rm dyn}=\alpha \xi_{\rm dyn}\approx\alpha / 4$. 
This is our basic dynamo prescription, except that we need some control
factors.

We want to make sure that the dynamo does not generate
too much poloidal magnetic field which would make the magnetic
field angle exceed the expected value (Eq.~\ref{e.dyn7}).
In three-dimensions, the dynamo is self-regulated, but
it has to be limited manually in axisymmetry. We 
therefore introduce a factor,
\be
\label{e.dyn85}
f_{\xi}={\rm max}\left(\frac{\xi_{\rm dyn}-\xi}{\xi_{\rm dyn}},0\right),
\ee
which turns the dynamo off in regions that already have sufficient 
poloidal magnetic field.

The dynamo results partly from magnetic buoyancy which 
bends toroidal magnetic field lines and allows them
to create poloidal field. Therefore, it is reasonable
to assume that the dynamo effect is antisymmetric with
respect to the equatorial plane. We impose this by 
introducing a factor,
\be
\label{e.dyn9}
f_{\rm eq}=\frac{R}{H}\frac{\pi/2-\theta}{\pi/2},
\ee
where $H$ and $R$ are the local density scale height and
radius, respectively, and $\theta$ is the polar coordinate.

The magnetic field saturates at the level discussed here
only in the turbulent body of a disk. In the corona, 
the dynamo may not be as effective, and may lead to a completely
different saturation state. Therefore we limit the
sub-grid dynamo model to the interior of the disk, by 
introducing another scaling factor,
\be
\label{e.dyn10}
f_{\theta}={\rm max}\left(1-\left(\frac {\pi/2-\theta}{\Theta_H}\right)^2,0\right),
\ee
where $\Theta_H$ is the polar angle corresponding to
the density scaleheight calculated after each time step according
to Eq.~\ref{eq.scaleheight}.

Finally, we wish to eliminate the dynamo within the plunging region
where the flow is more laminar then turbulent. We achieve this
through another factor,
\be
\label{e.dyn11}
f_{R}=\frac{1}{1+\exp\left(-2.94\frac{R-R_{\rm ISCO}}{0.1R_{\rm ISCO}}\right)},
\ee
which smoothly goes from $f_{R}=1$ for $R\gg R_{\rm ISCO}$ to $f_{R}=0$ close
to the BH horizon.

Applying the factors introduced above into Eq.~\ref{e.dyn8} we get 
the ultimate formula for orthonormal coordinates,
\be
\label{e.dyn12}
\pder{A_{\hat \varphi}}{t}=\alpha_{\rm dyn}\Omega_{\rm K} R B_{\hat \varphi}f_Rf_\theta f_\xi f_{\rm eq}.
\ee
This can be generalized to non-orthonormal coordinates
by,
\be
\label{e.dyn13}
\pder{A_\varphi}{t}=\alpha_{\rm dyn}\Omega_{\rm K} R g_{\phi \phi}B^{\varphi}f_Rf_\theta f_\xi f_{\rm eq},
\ee
where for simplicity we considered only diagonal terms of the metric.
The above formula is applied in a finite difference form (Eq.~\ref{eq.dynamoaphi})
after each time step to generate a dynamo-induced purely toroidal vector
potential,  which provides a weak correction to the poloidal magnetic field, which is
then super-imposed on top of the preexisting field.

\subsection{$\Omega$-effect}

The mechanism described in the previous
section constantly injects poloidal magnetic field
in regions where the magnetic field angle (roughly equal to the
ratio of poloidal to toroidal components) is too
low. Radial magnetic field created this way
is then sheared by the differential rotation creating
extra toroidal component. In axisymmetry there is
no mechanism to suppress this
mechanism, and the gas quickly becomes over-magnetized.
In 3D the magnetic pressure saturates
roughly at $10\%$ of the total pressure \citep[Eq.~\ref{e.betaprime}, ][]{blackman+08,hirose09a,jiang+13}.
To enforce this condition we 
damp the lab-frame toroidal component of the magnetic field on the
dynamical timescale according to,
\be
\label{eq.dynamodamp2}
\frac{dB^\varphi}{dt}=-\alpha_{\rm damp}\, \Omega_{\rm K} B^\varphi f_R f_\theta f_{\beta'},
\ee
where $f_R$ (Eq.~\ref{e.dyn11}) and $f_\theta$ (Eq.~\ref{e.dyn10}) ensure that the damping occurs only outside the ISCO and within one scale-height, and the factor $f_{\beta'}$,
\be
\label{eq.dynamodamp3}
f_{\beta'}={\rm max}\left(\frac{\beta'-\beta'_{\rm damp}}{\beta'_{\rm damp}},0\right),
\ee
restricts the damping to over-magnetized
regions ($\beta'>\beta'_{\rm damp}=0.1$). 

\subsection{Validation}

To evaluate the effectiveness of the sub-grid dynamo model,
and to compare it with 3D simulations, we performed
a test non-radiative, global, GRMHD simulation set up
in a similar way to the $a_*=0.0$ SANE model from
\cite{narayan+12}. We used exactly the same initial 
torus as in the reference 3D simulation.
We threaded the torus with multiple quadrupolar loops of a weak 
magnetic field, resembling the initial state for the
corresponding radiative runs discussed in the main part of the paper.
We used a grid of 256x256 points with  cells spaced 
logarithmically in radius and uniformly in 
 polar angle. The adopted vertical resolution was
a factor of two higher than in \cite{narayan+12}, because
the reference 3D run was only marginally resolved.
We adopted the same dynamo parameters as for the
radiative runs, i.e., $\alpha_{\rm dyn}=0.05$,
 $\xi_{\rm dyn}=0.25$, $\alpha_{\rm damp}=1.0$ and $\beta'_{\rm damp}=0.1$.

\begin{figure}
\includegraphics[width=1.075\columnwidth]{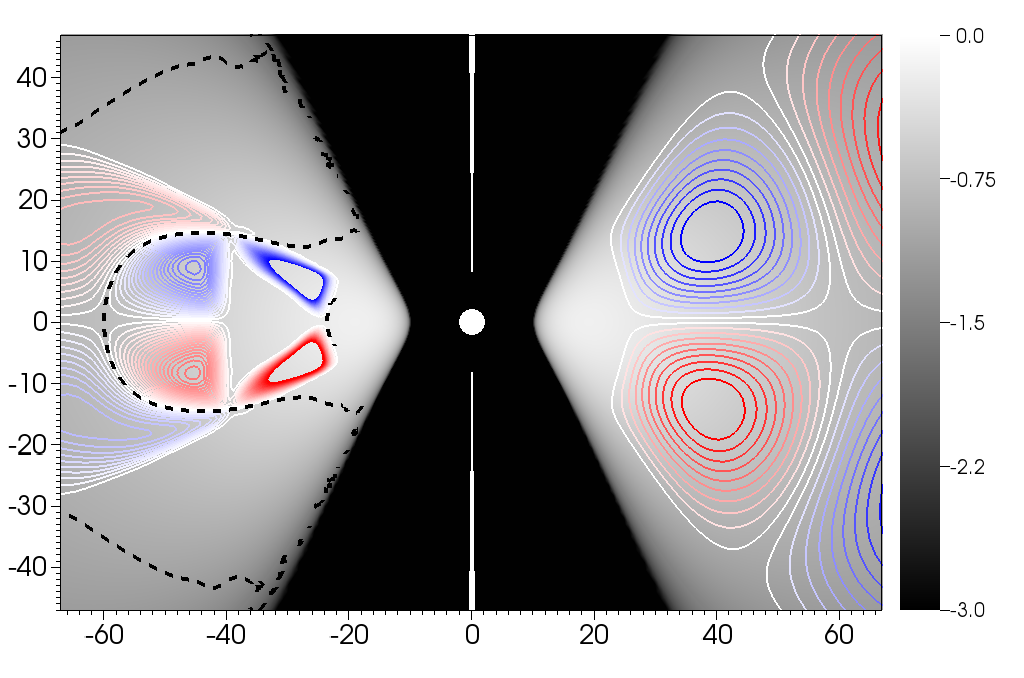}
\includegraphics[width=1.075\columnwidth]{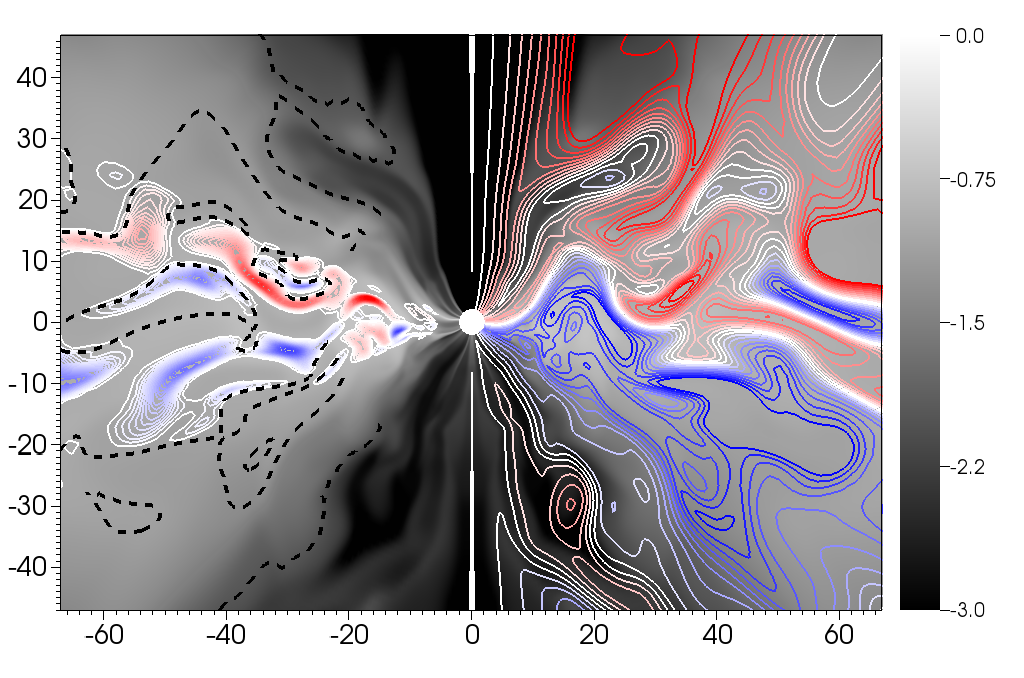}
\caption{
Snapshots of the test, non-radiative GRMHD simulations, showing
the poloidal magnetic field (right) and the dynamo-generated
correction (left panel). Contours show the magnetic field lines.
Red and blue correspond to clockwise and counter-clockwise
loops, respectively.
Grey colors show the logarithm of density.
The top and bottom panels corresponds to an early stage of the simulation
($t=100$), and a late one ($t=100,000$), respectively.
The  dashed contour in the left panels shows where $R B^\varphi$  
changes sign.
}
\label{f.dynamos}
\end{figure}

Figure~\ref{f.dynamos} shows the dynamo-generated
correction to the poloidal magnetic field at  early
(top), and late (bottom panel) times in the simulation.
At $t=100$ (top panel), the MRI has not yet developed
and the poloidal magnetic field (right panel) follows
the initial configuration. However, although initially 
there was no azimuthal magnetic field, differential 
rotation quickly stretches the radial component and 
generates toroidal field, $B^\varphi$, which triggers
the dynamo action. The left panel shows the dynamo
correction (contour levels are arbitrary) which was obtained by
taking the curl of the dynamo-generated vector potential $A_\varphi$
(Eq.~\ref{e.dyn13}) which follows $\propto R B^\phi$. 
One may expect flipping polarity
of the resulting dynamo field where this
quantity changes sign. This is indeed the case, 
as the dashed contours in the left panel of Figure~\ref{f.dynamos}
show. A similar flip takes place at the equatorial plane
because of the assumed antisymmetry (Eq.~\ref{e.dyn9}).
The vertical extent of the 
region affected by the dynamo  is limited by
the density scaleheight through Eq.~\ref{e.dyn10}.

The bottom panel of Figure~\ref{f.dynamos} shows
an advanced stage of the test simulation ($t=100,000$).
The MRI-driven turbulence
is still evident despite the very late time
 (for comparison turbulence in the same disk evolved without 
the sub-grid dynamo decays after $t\approx 5,000$).
Turbulent disk structure implies turbulent distribution
of $B^\varphi$ which results in more chaotic dynamo contribution,
as seen in the left panel. Once again, the poloidal loops flip signs
where $RB^\varphi=0$ (dashed contours), and at the equatorial plane.

\begin{figure}
\includegraphics[width=1.0\columnwidth]{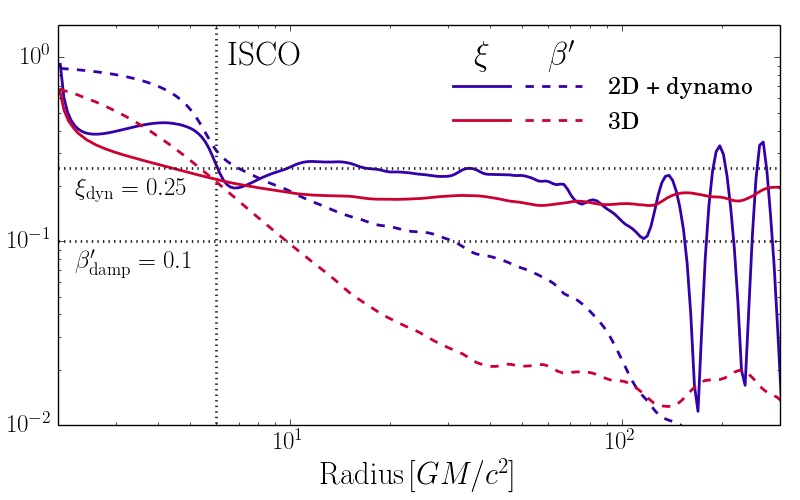}
\caption{
Magnetic field angle $\xi$ (Eq.~\ref{eq.bangle}, solid lines), and 
 magnetic to total pressure ratio $\beta'$ (dashed lines) 
for a test axisymmetric GRMHD run with dynamo (blue lines), and a full 3D simulation \citep[][, red lines]{narayan+12}.
The horizontal dotted lines show the target values of $\xi$ and $\beta'$ 
used by the sub-grid dynamo model. The vertical line denotes the radius of the ISCO.}
\label{f.dynamobetatheta}
\end{figure}

To obtain the mean properties of the magnetic field, we averaged the
disk solution over time $t=50,000\div 100,000$.  In
Figure~\ref{f.dynamobetatheta} we plot the magnetic field angle $\xi$
(solid lines), and the magnetic to total pressure ratio $\beta'$
(dashed lines) for our test simulation, and, for comparison, for the
3D reference run. In the test simulation, the magnetic field angle
settled down on the prescribed value $\xi_{\rm dyn}=0.25$ for the
whole converged region ($R\lesssim 75$) outside the marginally stable orbit. The
magnetic field angle inside the ISCO, where the dynamo model is not
applied (Eq.~\ref{e.dyn11}), departs from $\xi_{\rm dyn}=0.25$ and
increases, although non-monotonically, towards the horizon.  The mean
magnetic field angle obtained in the 3D simulation is closer to $0.2$
and smoothly increases towards the horizon. The two profiles are in
qualitative agreement. We suspect that the 3D run gave an angle of 0.2 
rather than the more accurate valueo of
$0.25$ \citep{sorathia+12},  because of insufficient resolution.

The magnetic to total pressure ratio in our test axisymmetric
simulation is close to the prescribed value $\beta'_{\rm damp}=0.1$
near the radius of convergence ($R\approx 75$) and slowly increases
inward, reaching $\beta'\approx0.3$ at the ISCO. Inside that critical
radius, it grows rapidly towards the horizon as the gas in the
plunging region is highly magnetized. The departure of the pressure
ratio from the target value $\beta'_{\rm damp}=0.1$ results from the
fact that the radial velocity in  a radiatively innefficient flow
is large enough not to give the damping prescription
(Eq.~\ref{eq.dynamodamp2}) enough time to balance the shearing-induced
generation of the toroidal magnetic field. A similar increase
in magnetization is clear also in the 3D run. However, the value of 
the pressure ratio in the 3D simulation was significantly 
lower ($\beta'=0.02$ for $R\gtrsim 30$),  is inconsistent with
results from local shearing sheet studies \citep[e.g.,][]{blackman+08,jiang+13}.
Once again, we suspect this is a result of inadequate resolution.

\section{Radiative viscosity for M1 closure}
\label{ap.viscosity}

The radiative closure scheme adopted in this work, the M1 closure,
limits the set of possible angular distributions of the radiation.
M1 assumes that the specific intensity in the lab frame can be 
obtained by boosting an isotropic distribution. This is a strong 
assumption and, obviously, often will not be satisfied. 
The simplest case in which M1 closure fails, is
a problem consisting of two discrete, isotropic sources of light,
emitting into an optically thin medium.
M1 closure allows for specific intensities 
boosted only along one direction -- the direction of the net flux (while the photons want
to cross each other and propagate in two distinct directions).
The result is that M1 causes an unphysical pile up of photons where the two beams cross.

Fortunately, accretion disks almost never have distinct point sources of
radiation where the radiation field can clash with other point sources. 
Photons are generated at the photosphere, which is a continuous extended source 
(though it may be turbulent and clumpy). Therefore, M1 closure is expected to work reasonably well for disks.

However, M1 closure has another flaw which may be important 
for BH accretion disks. As discussed in the Appendix of
\cite{sadowski+koral}, photons emitted from rotating disk,
when approaching the polar axis, mix in a way that overestimates
their angular momentum and prevents radiation from reaching the axis, 
creating a shock-like discontinuity in the radiative field just off
the axis. In simulations, this feature appears intermittently due to the 
turbulent nature of the accretion flow, and is more pronounced 
for thin disks.

In this paper we overcome\footnote{To rephrase Polish journalist Stefan Kisielewski commenting on socialism: ``M1 closure heroically
  overcomes difficulties unknown to any other closure''.} this
flaw by introducing an artificial viscosity in the radiation field
 which diffuses the shock and drives the radiation field towards a 
 more realistic distribution. 

In hydrodynamics, the viscous contribution
to the gas stress-energy tensor is often calculated as the product of
a dynamic viscosity coefficient and the shear tensor which describes the
rate of shear of fluid velocities \citep{landau+fluid}.
In the covariant formulation of the
M1 closure used in this paper, the radiation field is described 
using the velocity of the radiation rest frame, $u^\mu_{R}$ and its
energy density in that frame, $E_R$. The correspondence between 
the description of the gas and radiation, makes it possible
to apply a similar formalism for viscosity in context of the radiation
field by using the radiative rest frame velocities to calculate 
the (radiative) shear tensor, and to use the radiative energy density
to calculate the corresponding viscosity coefficient.

We therefore  introduce a viscous correction to the radiative
stress energy tensor,
\be
\label{e.Rvisc}
R^{\mu \nu}_{\rm visc}=-2\nu E_{ R} \sigma^{\mu \nu}
\ee
where $\nu$ is the viscosity coefficient (defined below), 
$E_R$ is the radiative energy density in the radiation rest
frame, and $\sigma^{\mu \nu}$ is the shear tensor, calculated 
from the radiation rest frames velocity field $u^\mu_{\rm R}$, through,
\be
\label{e.shear}
\sigma^{\mu \nu}=\frac12\left(u^\mu_{{\rm R};\alpha}P^{\alpha\nu}+u^\nu_{{\rm R};\alpha}P^{\alpha\mu}\right)-\frac13P^{\mu\nu},
\ee
where $P^{\alpha \beta}=g^{\alpha\beta}+u_{\rm R}^\alpha u_{\rm R}^\beta$ is the projection tensor.
The derivatives are calculated using radiative
velocities from the previous time step, and the time derivatives
are set to zero for stability \citep[compare][]{duez+04}.
The viscous coefficient $\nu$ is set to some fraction ($\alpha_{\rm
  rv}$) of
the mean free path of photons, $\lambda$,
\be
\label{e.viscnu}
\nu=\alpha_{\rm rv}\lambda,
\ee
where we estimate the mean free path as the inverse of the mean
opacities, but limited in the optically thin limit to the local
radial coordinate $R$, 
\be
\label{e.viscmfp}
\lambda={\rm min}\left(R,\frac1{\rho(\kappa_{\rm a}+\kappa_{\rm es})}\right).
\ee
The radiative viscosity is supressed in optically thick
regions because $\lambda$ becomes very small. 

The viscous correction to the radiative stress-energy tensor
(Eq.~\ref{e.Rvisc})
is applied only when calculating fluxes at the cell faces
(Eq.~\ref{eq.cons3_3}). Because we do not modify the algorithm,
and apply the advective operator explicitly, we need to make sure
that the viscous correction does not violate the usual Courant 
stability criterion
\citep{numericalrecipes}. For this purpose, we limit $\nu$
(Eq.~\ref{e.viscnu})
with,
\be
\label{e.viscnulim}
\nu_{\rm lim}=\frac{\Delta x_{\rm min}}{2\Delta t},
\ee
where $\Delta x_{\rm min}$ is the length of the shortest cell edge, and
$\Delta t$ is the timestep.

The magnitude of the radiative viscosity is controlled
by the parameter  $\alpha_{\rm rv}$. To choose the right
value we compare the synthetic radiation field generated by 
a radiating slab resembling a thin accretion disk, with 
the analytical solution obtained in the Newtonian approximation following 
\cite{tajimafukue-98}.
The latter is given by,
\be
\label{e.anathin}
\epsilon(R,z)\propto\int_0^\infty\int_0^{2\pi}\left(1-\sqrt{\frac 6{R_{\rm d}}}\right)R_{\rm d}^{-3}
\frac{I_{\rm M1}(\mu)}{(1+z_{\rm Dop})^4}\,{\rm d}\Omega,
\ee
where $R_{\rm d}$ is the radius at the equatorial plane, $I_{\rm
  M1}(\mu)$ is the angular distribution of photons at the disk plane
with
respect to angle $\mu$, $z_{\rm Dop}$ is the Doppler factor due to 
the orbital motion in the disk plane, $v^\varphi=R^{-1/2}$,
and ${\rm d}\Omega$ is an element of solid angle. 

We set up a corresponding problem in \texttt{KORAL} on a grid in spherical 
coordinates (no BH in the center), with radial coordinate spaced
logarithmically. The opacities were set to zero, the radiative energy
density at the lower, emitting boundary was set to,
\be
\label{e.m1thinbc}
\hat E\propto \left(1-\sqrt{\frac 6{R}}\right)R^{-3},
\ee
for $R>6$, and the orthonormal radiative flux in the comoving frame was set to,
\be
\label{e.m1thinbcflux}
 F^{\hat\theta}=-0.5 \hat E,
\ee
as appropriate for radiation escaping from optically thick slab. The
corresponding
specific intensity in the lab frame may be obtained by boosting the isotropic distribution 
with radiation rest frame velocity $\beta\approx0.394$, to get \citep{rybicki-book},
\be
\label{e.m1beam}
I_{\rm M1}(\mu,\beta)\propto \frac{(1-\beta^2)^2}{(1-\beta\mu)^4},
\ee
which describes the effective limb darkening of radiation from disk surface, 
and is used in Eq.~\ref{e.anathin}. This form of limb darkening
is implied by the intrinsic M1 assumption about the form of specific
intensities,
and, by accident, matches quite well with limb darkening induced
from vertical structure of emitting disk \citep{chandra-50}.

\begin{figure}
\includegraphics[width=.45\columnwidth]{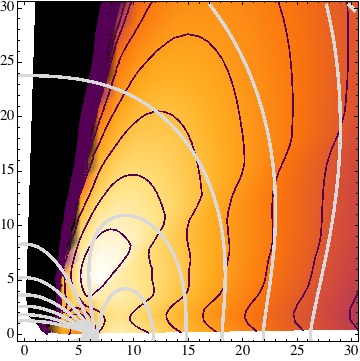}\hspace{.5cm}
\includegraphics[width=.45\columnwidth]{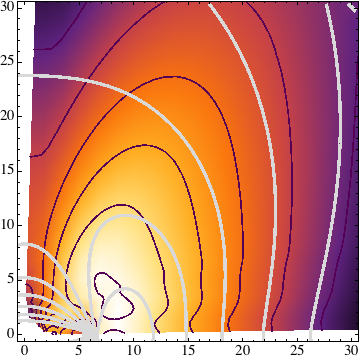}\vspace{.5cm}\\
\includegraphics[width=.45\columnwidth]{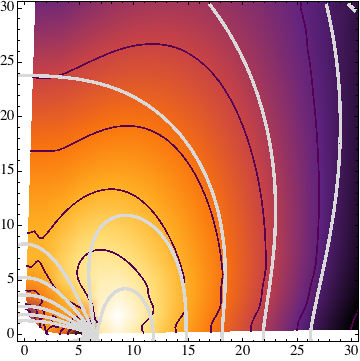}\hspace{.5cm}
\includegraphics[width=.45\columnwidth]{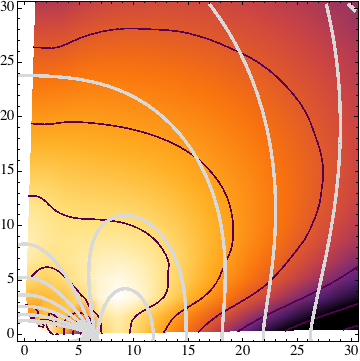}
\caption{Comparison of the analytical radiation field (Eq.~\ref{e.anathin}, light gray contours) with
the numerical one (colors and dark contours) calculated using the M1 scheme and the radiative
viscosity with $\alpha_{\rm rv}=0.0$, $0.05$ (top panels, left to right),$0.1$, and $0.5$ (bottom panels).
}
\label{f.radvisc}
\end{figure}

\begin{figure}
\includegraphics[width=1.0\columnwidth]{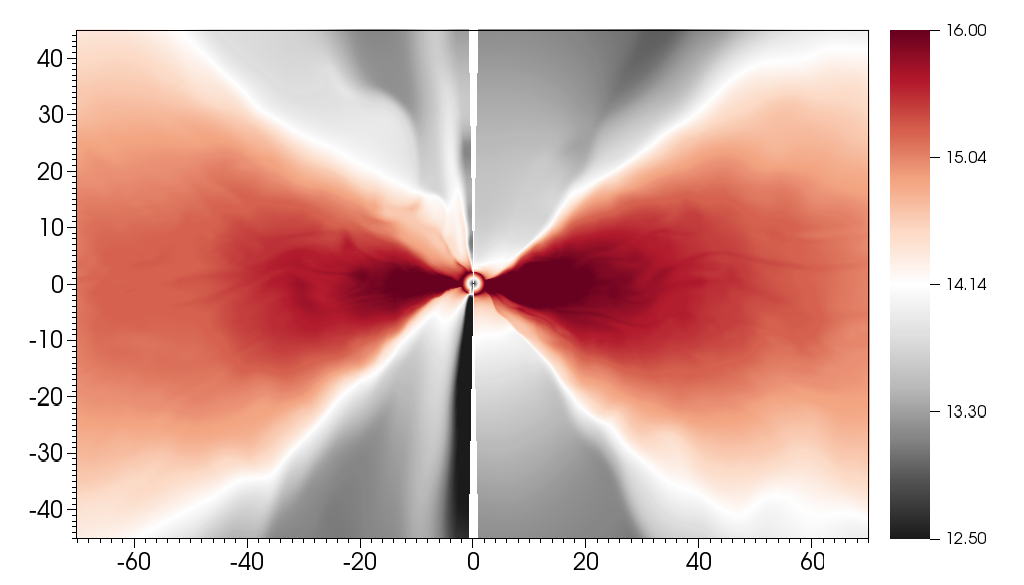}
\caption{Comparison of the
radiative energy distribution for runs without (left) and with (right
panel, \texttt{r300a0}) radiative viscosity at $t=65,000$.}
\label{f.viscnovisc}
\end{figure}

Figure~\ref{f.radvisc} compares the analytical and numerical
solutions.
White contours show the radiative energy distribution obtained
from Eq.~\ref{e.anathin}. Colors and black contours show
the numerical solution obtained with \texttt{KORAL} for $\alpha_{\rm rv}=0$
(first panel, no viscosity), $0.05$, $0.1$, and $0.5$ (lower right
panel). The numerical solution with no viscosity exhibits an unphysical funnel
near the axis that is devoid of radiation. This feature disappears
once radiative viscosity is included. Out of the three 
values of $\alpha_{\rm rv}$ considered, $\alpha_{\rm rv}=0.1$ provides
the best match to the analytical solution, and is the value adopted
in the simulations described in the main part of this paper. 
Figure~\ref{f.viscnovisc} shows a snapshot of the \texttt{r300a0}
simulation
at $t=65,000$ (right), and a corresponding snapshot from
an otherwise identical simulation, but with no radiative viscosity (left panel). The 
latter clearly shows an empty, unphysical funnel near the axis (especially in the lower half). The empty region appears intermittently.

\end{document}